\documentclass{aa}

\usepackage{graphicx}
\usepackage{natbib}
\usepackage{longtable}
\usepackage{booktabs}
\usepackage{rotating}
\usepackage{txfonts}
\usepackage{textcomp}
\setlongtables

\begin{document}

\title{The bipolar jet of the symbiotic star \object{R~Aquarii}: A study of its morphology using the high-resolution HST WFC3/UVIS camera.}

\author{Stanislav Melnikov \inst{1,3} \and Matthias Stute \inst{2} \and
  Jochen Eisl\"offel \inst{1}}

\institute{Th\"uringer Landessternwarte Tautenburg, Sternwarte 5, 07778 Tautenburg, Germany  \and 
            Simcorp GmbH, Justus-von-Liebig-Strasse 1, 61352 Bad Homburg, Germany \and
           Ulugh Beg Astronomical Institute, Astronomical str. 33, 700052 Tashkent, Uzbekistan}

\offprints{Stanislav Melnikov,
\email{smeln2005@gmail.com}}

\date{Received / Accepted}

\abstract{\object{R~Aqr} is a symbiotic binary system consisting of a Mira variable with a pulsation period of 387 days and a hot companion which  is
presumably a white dwarf with an accretion disk. This binary system is the source of a prominent bipolar gaseous outflow.}
{We use high spatial resolution and sensitive images from the \textit{Hubble Space Telescope} (HST) to identify and investigate the different 
structural components that form the complex morphology of the R~Aqr jet.
}
{We present new high-resolution HST WFC3/UVIS narrow-band images of the R~Aqr jet obtained in 2013/14 using the [OIII]$\lambda$5007, 
[OI]$\lambda$6300, [NII]$\lambda$6583, and H$\alpha$ emission lines. These images also allow us to produce detailed maps of the jet flow in several 
line ratios such as [OIII]$\lambda$5007/[OI]$\lambda$6300 and [NII]$\lambda$6583/[OI]$\lambda$6300 which are sensitive to the outflow temperature and 
its hydrogen ionisation fraction. The new emission maps together with archival HST data are used to derive and analyse the proper motion of prominent 
emitting features which can be traced over 20 years with  the HST observations.
}
{The images reveal the fine gas structure of the jet out to distances of a few tens of arcseconds from the central region, as well as in the innermost 
region, within a few arcseconds around the stellar source. They reveal for the first time the straight, highly collimated jet component which can be 
traced to up to $\sim900$ AU in the NE direction. Images in [OIII]$\lambda$5007, [OI]$\lambda$6300, and [NII]$\lambda$6583 clearly show a helical 
pattern in the jet beams which may derive from the small-scale precession of the jet. The highly collimated jet is accompanied by a wide opening angle 
outflow which is filled by low excitation gas. The position angles of the jet structures as well as their opening angles are calculated. Our 
measurements of the proper motions of some prominent emission knots confirm the scenario of gas acceleration during the propagation of the outflow. 
Finally, we produce several detailed line ratio maps which present a mosaic combined from the large field and the PSF-subtracted inner region.
}
{\textrm{The high signal-to-noise HST WFC3/UVIS images provide powerful tools for the study of the jet morphology and also bring detailed information 
about the physical jet gas conditions.} The simultaneous observations of [OIII], [OI], [NII], and [SII] would allow us to measure basic parameters 
of the ionised gas in the R~Aqr outflow such as electron density, electron temperature and hydrogen ionisation fraction, and compare them with 
other stellar jets.
}

\titlerunning{HST WFC3 observations of the R~Aqr jet}
\maketitle
\section{Introduction}
\label{intro}

The symbiotic binary system \object{R~Aqr} is one of the brightest ($V=7.7$) and nearest objects of this class \citep[D$\sim$220 pc, ][]{Min14}. 
This interacting binary consists of a Mira-like variable (Sp.Type M7) with a pulsation period of 387 days \citep{Mer35} and a secondary 
companion resolved recently by \citet{Sch17}. This object is believed to be a white dwarf (WD) surrounded by a 
gaseous disk which is formed from stellar wind material flowing from its red giant companion. The distance between the companions (semi-major axis) 
is estimated to be $\sim15$ AU and an orbital period of $\sim$44 yr has been derived \citep{Grom09}. The most striking feature of this system of 
evolved stars is a prominent bipolar gaseous outflow extending in a NE--SW direction. This bright outflow is visible in various spectral features --- 
optical hydrogen lines, for example, in H$\alpha$, optical forbidden lines, as well as at radio-, UV, and X-ray wavelengths.

Based on observations from X-ray to radio wavelengths, \citet{Burg92} proposed a model of R~Aqr in which the binary system is composed of a M7 Mira 
with a mass of $\sim1.5 M_\odot$ ($T_*\approx2800$) and a hot companion (a WD or a subdwarf) with $T_*\approx40 000$. \citet{Burg92} concluded that 
the origin of the jet could be a  stellar wind blowing out with a high velocity. On the other hand, other mechanisms, such as an accretion disk with a 
strong magnetic field, or the collision of two stellar winds cannot be excluded \citep{Burg92}. The nebular structures around R~Aqr present striking 
similarities with that of another symbiotic star, \object{HM Sge} \citep{Solf84,Solf85} and therefore, it may be a common phenomenon in evolved 
binaries. On the other hand, such jets are frequently observed in young stellar objects (YSO) in star forming regions and therefore, the existence of 
the bipolar jets and outflows around symbiotic stars may imply similar mechanisms for the origin of this phenomenon. One of the most popular scenarios 
for the YSO jet formation is based on magneto-hydrodynamic (MHD) models which invoke magneto-centrifugal acceleration along magnetic field lines 
threading an accretion disk \citep{Shu00,Fer04}. According to this scenario, the stellar source of the jet is surrounded by the disk and this outflow 
phenomena is intimately linked to the accretion-ejection processes.

Extensive studies in several emission lines revealed the complex structure of the R~Aqr outflow. Looking at the large scales, the R~Aqr outflow 
consists of an hourglass-like bipolar shell ($\sim40\arcsec$ in each direction) surrounded by an outer oval-shaped shell with a radius of about 
1$\arcmin$, located  in a plane approximately orthogonal to the outflow axis \citep{Solf85}. \citet{Solf85} concluded that the shells --- the 
bipolar shell and outer oval-shaped nebulae --- are formed due to outbursts with a time interval of  $\sim$450 yr, 180 and 640 years ago. 
\citet{Min14} estimated the expansion age of the bipolar shell to be $\sim$240 yr, corresponding to an outburst date about the year 1773. At small 
scales, high-resolution observations of R~Aqr reveal that an inner bipolar outflow is more collimated, extending over $\sim15\arcsec$ 
\citep{Solf85,Burg92}. \citet{Solf85} found that the symmetry axis of the hourglass-like outflow is inclined at about 18\textdegree\, with respect to 
the plane of the sky, but the axis of the inner outflow seems to be tilted to the basic axis of the hourglass-like shell. Comparison of radio- and 
[OIII]$\lambda$5007 emission showed a characteristic `S'-shape in the inner outflow close to the stellar system, which was interpreted as precession 
in the orbital plane of the binary \citep{Mic88,Hol93}. The same apparent morphology can be seen at larger distances as emission arcs filling the 
hourglass-like shell \citep{Solf85}. 

A new era of R Aqr jet studies began with using space missions such as \textit{Hubble Space Telescope} and \textit{Chandra}, which allowed us to 
significantly increase the spatial resolution and the flux sensitivity. The first HST images obtained in several emission lines in 1991--92 allowed 
\citet{Par94} to resolve a fine jet structure within $10\arcsec\times10\arcsec$ scale. These HST images revealed a series of knots with prominent 
parallel features lying transverse to the flow direction. \citet{Par94} concluded that the jet is already highly collimated within 15 AU from its 
source, that is, within the confines of the binary system. \citet{Hol97a} analysing ultraviolet HST images obtained over a two-year period found 
evidence for subarcsecond changes in the ultraviolet morphology ($\sim 2550$\ \AA) of the inner 5$\arcsec$ of the jet and derived that knot 
velocities increase with increasing distance from the central source, from $\sim40$ to $\sim 240$ km s$^{-1}$.

In addition to low-excitation forbidden lines such as [OI]$\lambda$6300 and [NII]$\lambda$6583, the R~Aqr outflow is also bright in some emission 
lines of high excitation, such as [OIII]$\lambda$5007 and O\,VI\,$\lambda$1032 \citep{Nic09}, for example. The [OIII]$\lambda$5007 line is also 
observable in some YSO jets, but is usually relatively faint and detected in a few bow shocks only. Moreover, \citet{Kel01} resolved the point source 
X-ray emission from R~Aqr \citep{Jura84} into a chain of X-ray knots demonstrating a super-soft spectrum from the jet beam. 

New high-resolution sensitive instruments allow us to investigate the structure of the R~Aqr outflow in more detail. In this paper, we present high-resolution HST narrow-band images of the R~Aqr jet and outflow obtained in several bright optical emission lines. 
The paper is focused on the study of the outflow morphology and organised as follows. In Section 2, we describe the HST WFC3/UVIS observations 
obtained in 2013/14. High-resolution emission maps for the outer and inner parts of the R~Aqr outflow derived from these data are presented in Section 
3. We are showing several two-line ratio maps for the jet and discuss the line ratio distribution in Section 4. In Section 5, we present the proper 
motions computed using the 2013/14 observations together with archival HST data. Finally, we discuss the structure of the R~Aqr outflow complex and 
summarise our results in Section 6.

\section{Observations and data analysis}
\label{obs}
Our high-resolution HST WFC3/UVIS images of the R~Aqr jet were taken during two HST orbits in 2013 (Oct 22) and 2014 (Oct 18). The narrow-band images 
in [OIII]$\lambda$5007, [OI]$\lambda$6300, [NII]$\lambda$6583, and H$\alpha$ were obtained as both short and long exposures. These UVIS images have a 
pixel scale of $0\arcsec.03962$. The detailed observing log in Table~1 gives the line and corresponding filter name, start UT and exposure time, mean 
JD time, and focus calculated with the HST focus model for the times of the observations. The computed HST focus is a parameter necessary for 
generating the instrumental WFC3 Point Spread Function (PSF) pattern for those times.

The basic reduction of the images (bias, flatfielding, etc.) was obtained within the standard HST pipeline reduction. Depending on the exposure time, 
the HST images have been cleaned from cosmic ray hits. The long-exposure images have been combined from several shorter exposure shots (dithering 
process) which allowed the pipeline software to identify cosmic ray hits and to remove them from the resulting images. The short-exposure images, 
however, consist of a single-shot image without dithering and therefore they were not cleaned from cosmics, hot pixels, or other detector artefacts 
after the pipeline reduction. These short-exposure images were cleaned from cosmic hits using a Laplacian algorithm of cosmic ray identification 
\citep{Dok01}.
\begin{table}
\centering
\caption{\label{log} Journal of observations.}
\begin{tabular}{l@{\hspace{2mm}}c@{\hspace{2mm}}c@{\hspace{2mm}}c@{\hspace{2mm}}r@{\hspace{2mm}}r}
\hline\hline
Line                    &   Filter & UT(start) & Exposure &   JD(mean) & Focus \\
(\AA)                   &          & (h min secs)  &  (secs)  &   (24... days)  & (metres) \\ \hline \\
\multicolumn{6}{l}{2013, Oct 22}\\
\hline
{[OIII]} $\lambda$ 5007 &   F502N  & 21 23 58  &   15     &  56587.8917 & 12.11 \\
{[OIII]} $\lambda$ 5007 &   F502N  & 21 26 30  &   1086   &  56587.9005 & 10.03 \\
{[OI]} $\lambda$ 6300   &   F631N  & 21 49 26  &   1086   &  56587.9164 & 6.85  \\
{[OI]} $\lambda$ 6300   &   F631N  & 21 56 33  &   15     &  56587.9560 & 12.08 \\ \hline 
\\
\multicolumn{6}{l}{2014, Oct 18}\\ 
\hline
H$\alpha$               &   F656N  & 12 39 00  &   70     &  56948.5271 & 6.25  \\
H$\alpha$               &   F656N  & 12 44 47  &   2058   &  56948.5446 & 2.24  \\
{[OIII]} $\lambda$ 5007 &   F502N  & 13 26 20  &   50     &  56948.5600 & -0.04 \\
{[NII]} $\lambda$ 6583  &   F658N  & 14 17 49  &   70     &  56948.5958 & 5.58  \\
{[NII]} $\lambda$ 6583  &   F658N  & 14 20 16  &   2085   &  56948.6110 & 2.22  \\
{[OI]} $\lambda$ 6300   &   F631N  & 15 02 11  &   24     &  56948.6265 & -0.10 \\
\hline
\end{tabular}
\end{table}

Using the zero-point given in the fits-headers of the images, we converted the instrumental intensity (counts) into absolute fluxes in units of erg 
cm$^{-2}$ s$^{-1}$ \AA$^{-1}$. R~Aqr is a very nearby object with only low interstellar extinction. According to \citet{Nic09}, the north-east 
(NE) jet has $E(B-V) = 0.1$ whereas the south-west (SW) jet is unabsorbed, that is, $E(B-V) \approx 0$. To correct the fluxes for extinction for the 
NE-jet, \citet{Nic09} used the extinction curve of \citet{Fitz99} with the value of $E(B-V)$ and a normal interstellar extinction law 
($R_V = 3.1$). Since there is such an $E(B-V)$-asymmetry, most of the extinction is probably circumstellar and it may deviate from the interstellar 
law. However, since $E(B-V)$ for the NE-lobe is not high, the difference will not be important and we also adopted the normal interstellar law.

Every WFC3/UVIS image contained a flux pattern super-imposed by the instrumental WFC3 PSF which makes it difficult to analyse the jet structures 
close to the central jet source. The peak PSF should in general coincide with the position of the stellar object. However, in the case of the emission 
lines which trace the radiating ambient gas, the PSF centre can be shifted with respect to the stellar position. The long-exposure images, which 
provide a good signal-to-noise ratio (S/N) and are obtained in bright emission lines such as H$\alpha$ or [OIII]$\lambda$5007, showed that their central 
emission profiles are over-saturated and artificial spikes are generated. Therefore, the PSF pattern of these lines is difficult to model. To avoid 
that, the images with short exposures, which prevented saturation close to the central emission sources, were obtained as well. These short-exposure 
images were used to remove the saturation from the long-exposure star profiles and reconstruct a reliable distribution of the central emission. As a 
result, the final combined images were suitable for more accurate modelling of the PSF pattern and removing this pattern from the long-exposures 
images.

To model this HST PSF pattern, we used the TinyTim software with the web interface available at the STSCI site. We computed the HST PSF maps around 
the central jet source with a size of $10\arcsec\times10\arcsec$ where this pattern was particularly strong and removed this pattern in an iterative 
process with the Lucy-Richardson deconvolution algorithm. The number of iterations to reproduce the best solution for the PSF-fitting was estimated 
empirically for each image by visual inspection. The best solution was adopted when the PSF-pattern was already faint enough, but the S/N
remained high. On average, the number of iterations ranges between 15 and 30. These $10\arcsec\times10\arcsec$ flux maps cleaned from the PSF 
were then used to produce the emission-line ratio maps. In turn, the $10\arcsec\times10\arcsec$ ratio maps have also been used to construct the 
large-field line-ratio maps cleaned from the PSF pattern.

For accurate spatial analysis we implemented a precise astrometrical calibration into the HST frames. Some uncertainty of the astrometry may come from the accuracy of the world coordinate system (WCS) embedded into the HST images. Examination of positions of several stars found around R~Aqr showed that the HST WCS was very accurate in all images during the 2013 observing run, whereas during the 2014 run the derived coordinates fluctuated around some mean position. The offsets for each image in 2014 were calculated and the WCS were modified accordingly.

As a last step of the astrometric reduction, we took into account the proper motion of R~Aqr. The WFC3 camera has a pixel scale of $0\farcs04$ and the 
proper motion (PM) of R~Aqr is $\mu_{R.A.}=0\farcs033$ $\mu_{DEC}=-0\farcs026$. Therefore, this PM can already influence results from observations separated by just one year. Therefore, we took the PM for the epoch difference between the 2013 and 2014 runs into account and modified the WCS of the 2014 images to the  epoch of the 2013 images. As a result, the position of the central stellar source has been shifted to the single epoch.

\section{Jet morphology}
\label{emis}

The reduced high-sensitivity and high-resolution HST WFC3/UVIS images reveal the complex structure of the R~Aqr jet in unprecedented detail and allow us to 
study its morphology on different scales, from the outer lobe at $\sim45\arcsec$ scale to the inner part of the outflow within $1\arcsec$. The [OI]$\lambda$6300 and [NII]$\lambda$6583 lines outline areas of low excitation radiation and are produced by radiative shock rather than by photo-ionisation from the highly ionising radiation of the WD. Unlike these forbidden lines forming mostly due to a single process, H$\alpha$ emission can be produced both by collisional excitation, arising at high temperatures (several 10$^4$ K) and moderate ionisation, and by recombination, mainly occurring at low temperatures (less than 6--7000 K). The [OIII] emission can be produced by both fast shocks \citep[$V>100$ km/s][]{Har87} and the photo-ionisation from the hot jet source.

\subsection{Narrow-band images, large-scale maps ($45\arcsec$)}

\begin{figure*}[t]
\centering
\begin{minipage}[c]{9.1cm}
\resizebox{\hsize}{!}{\includegraphics{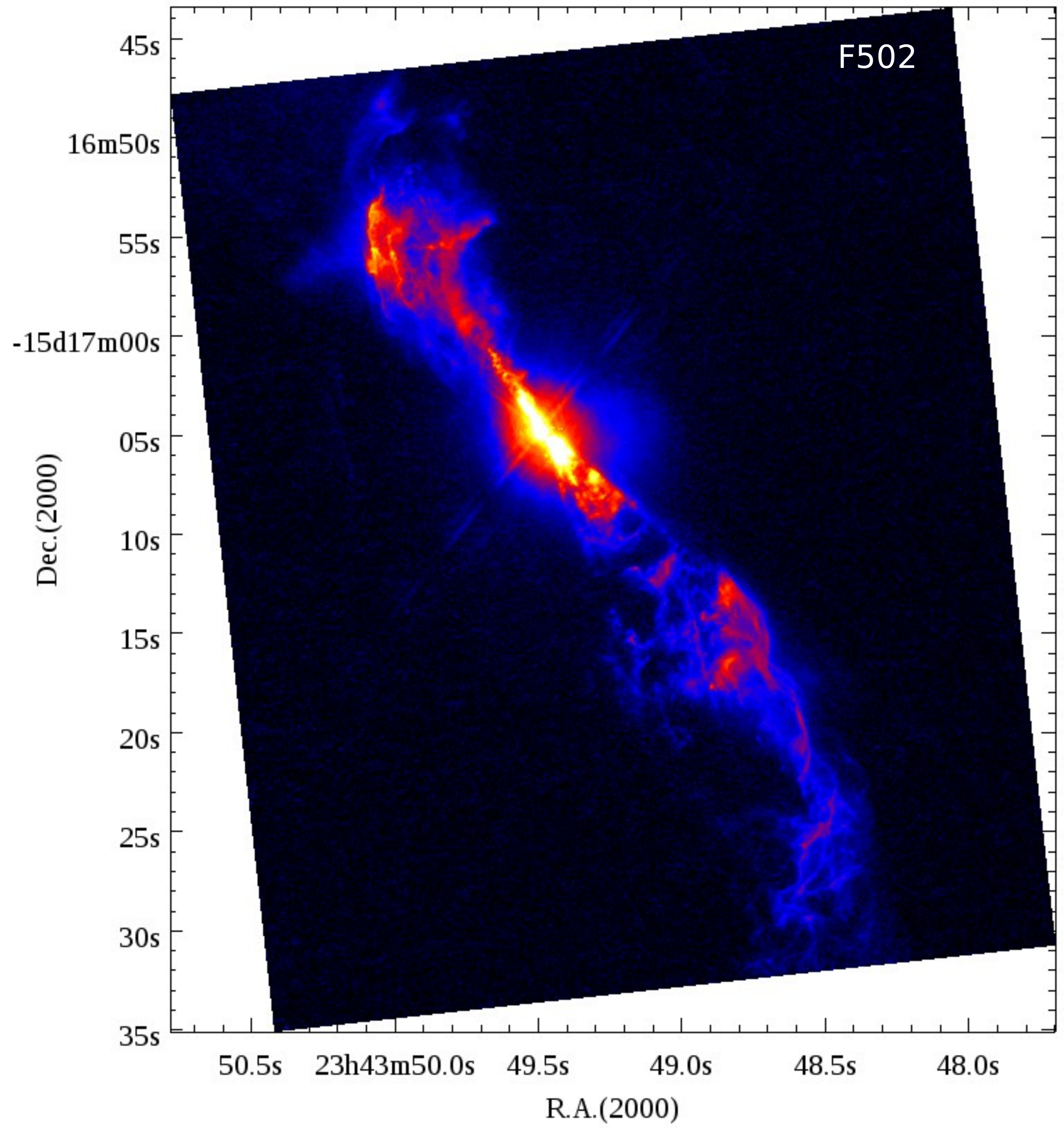}}
\end{minipage}
\begin{minipage}[c]{9.1cm}
\resizebox{\hsize}{!}{\includegraphics{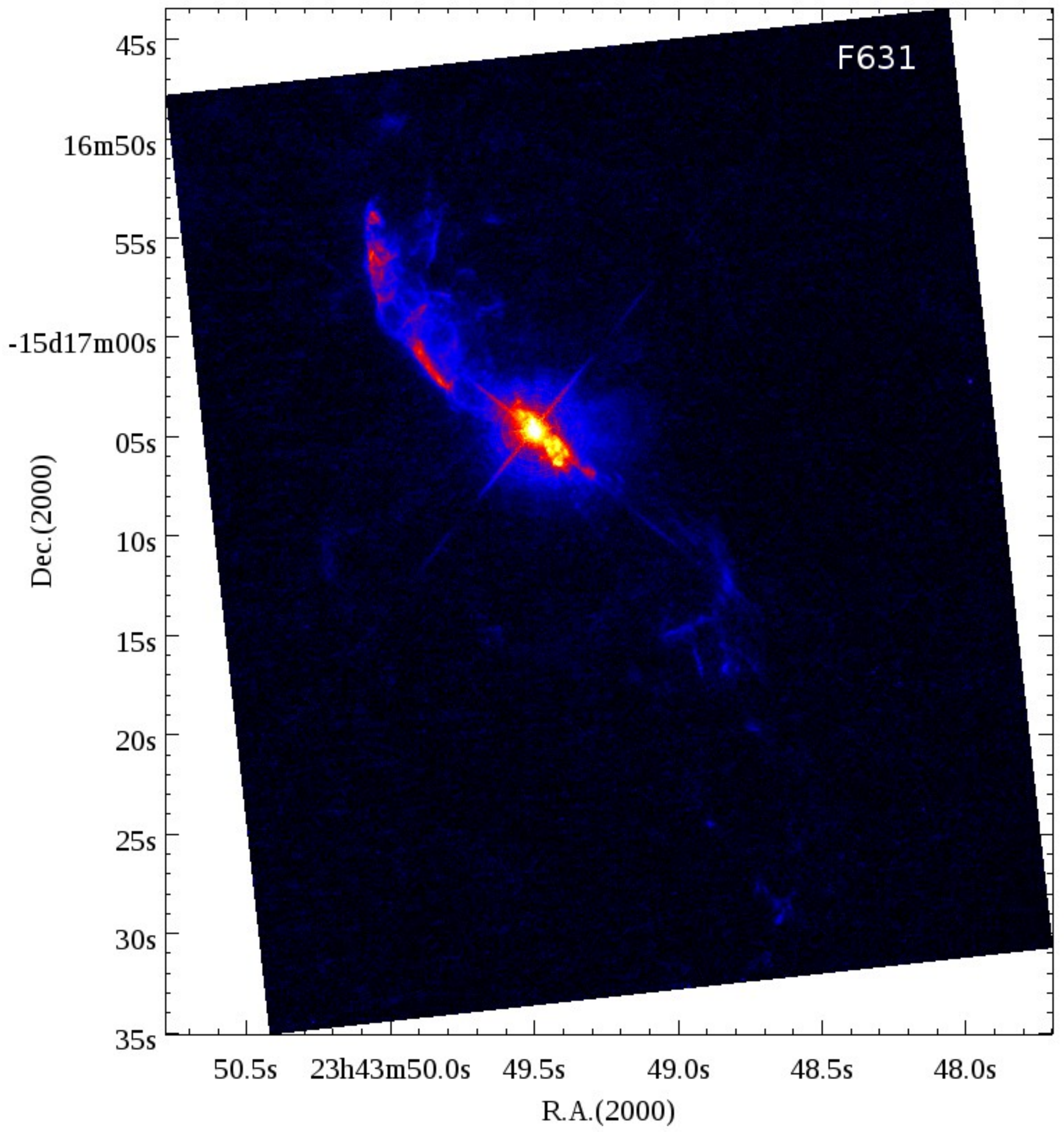}}
\end{minipage}

\begin{minipage}[c]{9.1cm}
\resizebox{\hsize}{!}{\includegraphics{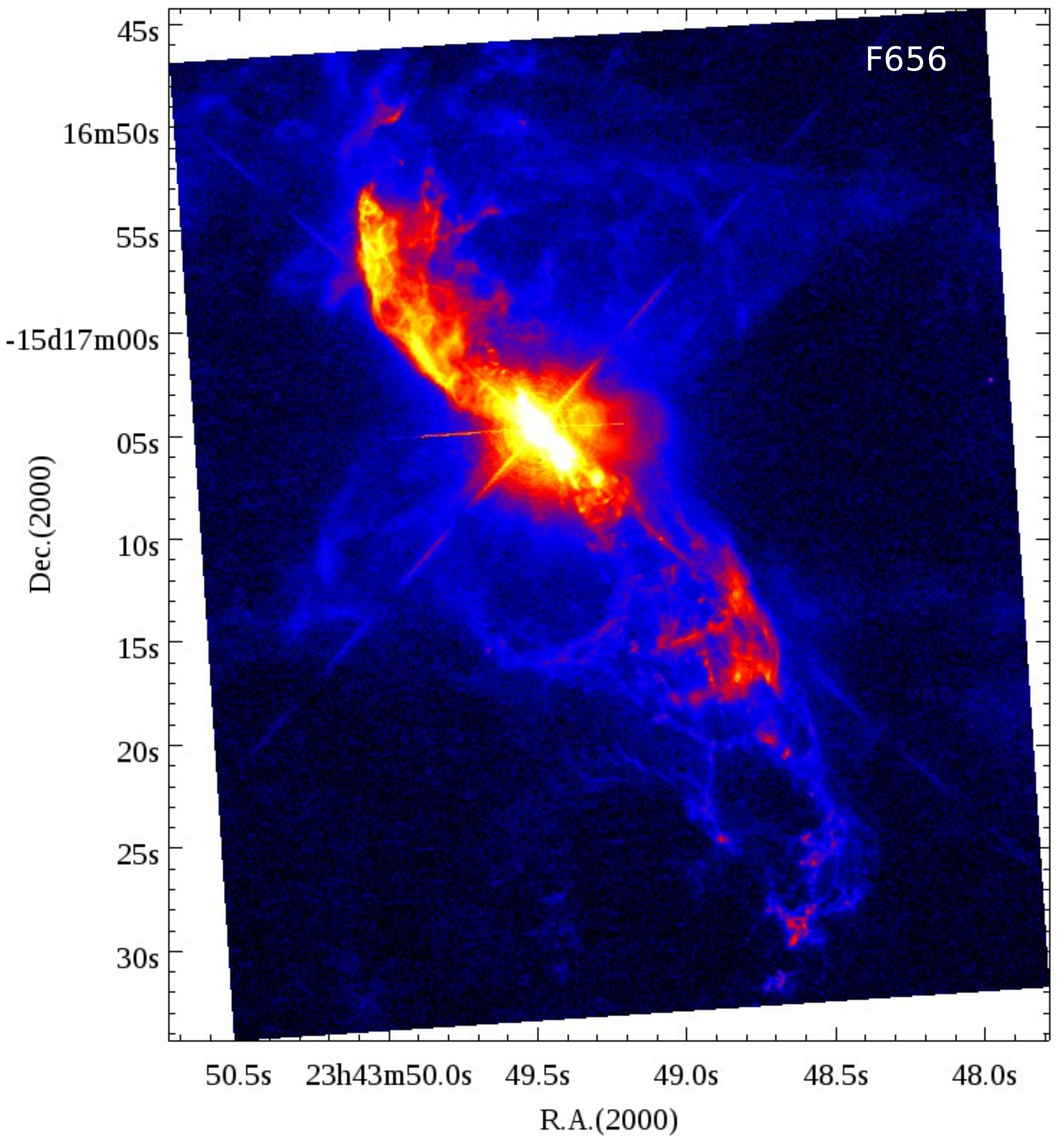}}
\end{minipage}
\begin{minipage}[c]{9.1cm}
\resizebox{\hsize}{!}{\includegraphics{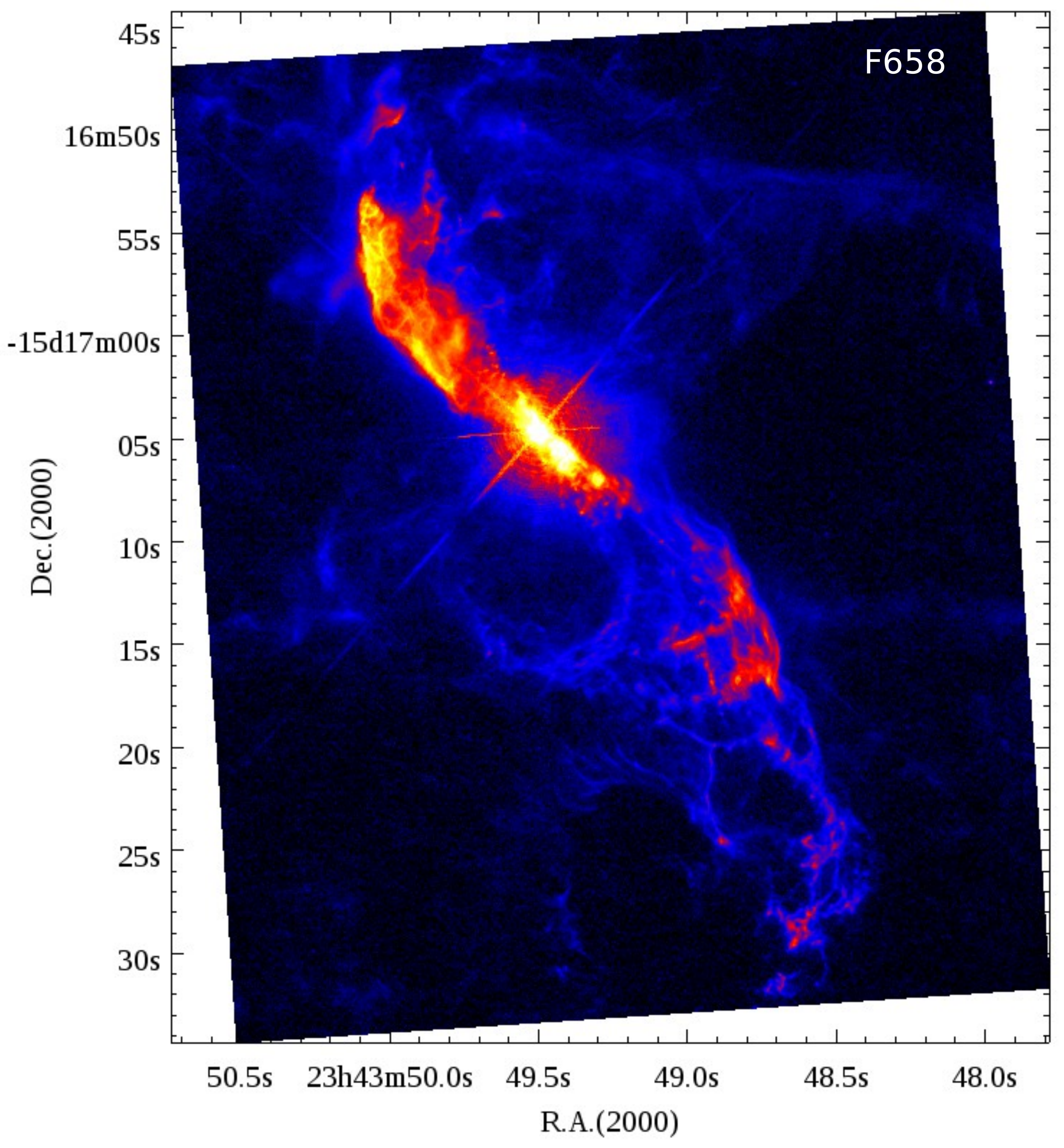}}
\end{minipage}

\begin{minipage}[c]{12cm}
\resizebox{\hsize}{!}{\includegraphics{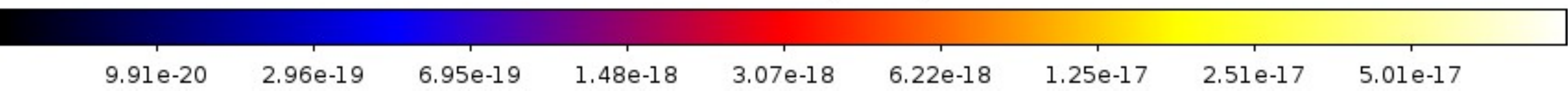}}
\end{minipage}
\caption{\label{flux_l} Large-scale HST images ($\sim45\arcsec\times45\arcsec$) taken through filters F502N ([O\,III]$\lambda$5007), F631N 
([OI]$\lambda$6300) during the 2013 observing run and the filters F656N (H$\alpha$) and F658N ([NII]$\lambda$6583) obtained in 2014. The colour scale 
for the fluxes ranges from 0 to $10^{-16}$ erg cm $^{-2}$ s$^{-1}$ \AA$^{-1}$. The H$\alpha$ map contains a roundish emission spot ($\sim 
2-3$ arcsec to the west of the central source) which is probably an "image ghost" (see also Fig.~\ref{flux_s}).}
\end{figure*}

The narrow-band flux-calibrated images illustrating the large-scale structure of the R~Aqr outflow taken in four emission lines --- 
[OIII]$\lambda$5007 (F502N), [OI]$\lambda$6300 (F631N), H$\alpha$ (F656N), and [NII]$\lambda$6583 (F658N) --- are shown in Fig.~\ref{flux_l}. For comparison, the emission in all the lines is plotted with the same flux scale, which is shown on the horizontal bar. The images resolve the 
large-scale outflow lobes elongated in NE and SW directions that is roughly aligned with the axis of the inner collimated jet, but does 
not exactly coincide with it (Fig.~\ref{flux_s}). We discuss the orientation of the emitting gaseous structures and the axis of the 
high-collimated jet in detail in the Discussion Section. 

One can see that the R~Aqr outflow shows a complex structure consisting of several arcs. A sequence of the emission arcs forms the curved lobes with a 
curvature pointing in opposite directions in the counterparts. Most prominent arcs are present in the SW lobe (bottom), but some can also be 
recognised in the NE counterpart. This shape is especially prominent in H$\alpha$ and [NII]. In [OIII], the arcs are less prominent than in the 
low-excitation line maps, but they are still clearly visible. The outflow in [OIII] appears to be straight and aligned mostly along the axis of the 
inner collimated jet. The [OIII] outflow also appears narrower than in low-excitation lines, with the bulk of emission located around the beam axis.

In contrast to [OIII], the H$\alpha$ and [NII] maps also reveal extended areas of faint emission (faint blue emission in the Fig.~\ref{flux_l}), which 
also shows a variety of patterns such as arcs, rims, and loops. One can see that this faint emission in H$\alpha$ and [NII] is visible even far from 
the outflow axis and spreads even to planes orthogonal to the jet propagation direction. [OI] is the faintest emission line among the studied species, 
but some emission arcs are readily seen. Due to the faintness of the [O\,I] emission, however, many structural details outside of the innermost 
outflow region are not visible. Apart from that, the morphology of [OI] resembles that of the other low-excitation lines, [NII] and H$\alpha$. It is 
evident, that the SW lobe in the low-excitation lines is fainter than the NE lobe, in contrast with the [OIII] emission, where the 
brightness of both lobes is rather similar. This brightness asymmetry visible in low-excitation emission reflects different excitation and temperature 
conditions existing in the opposite jet lobes, which is also often found in the YSO jets.

\subsection{Narrow-band images, $10\arcsec\times10\arcsec$ maps}

\begin{figure*}[t]
\centering
\begin{minipage}[c]{9.1cm}
\resizebox{\hsize}{!}{\includegraphics{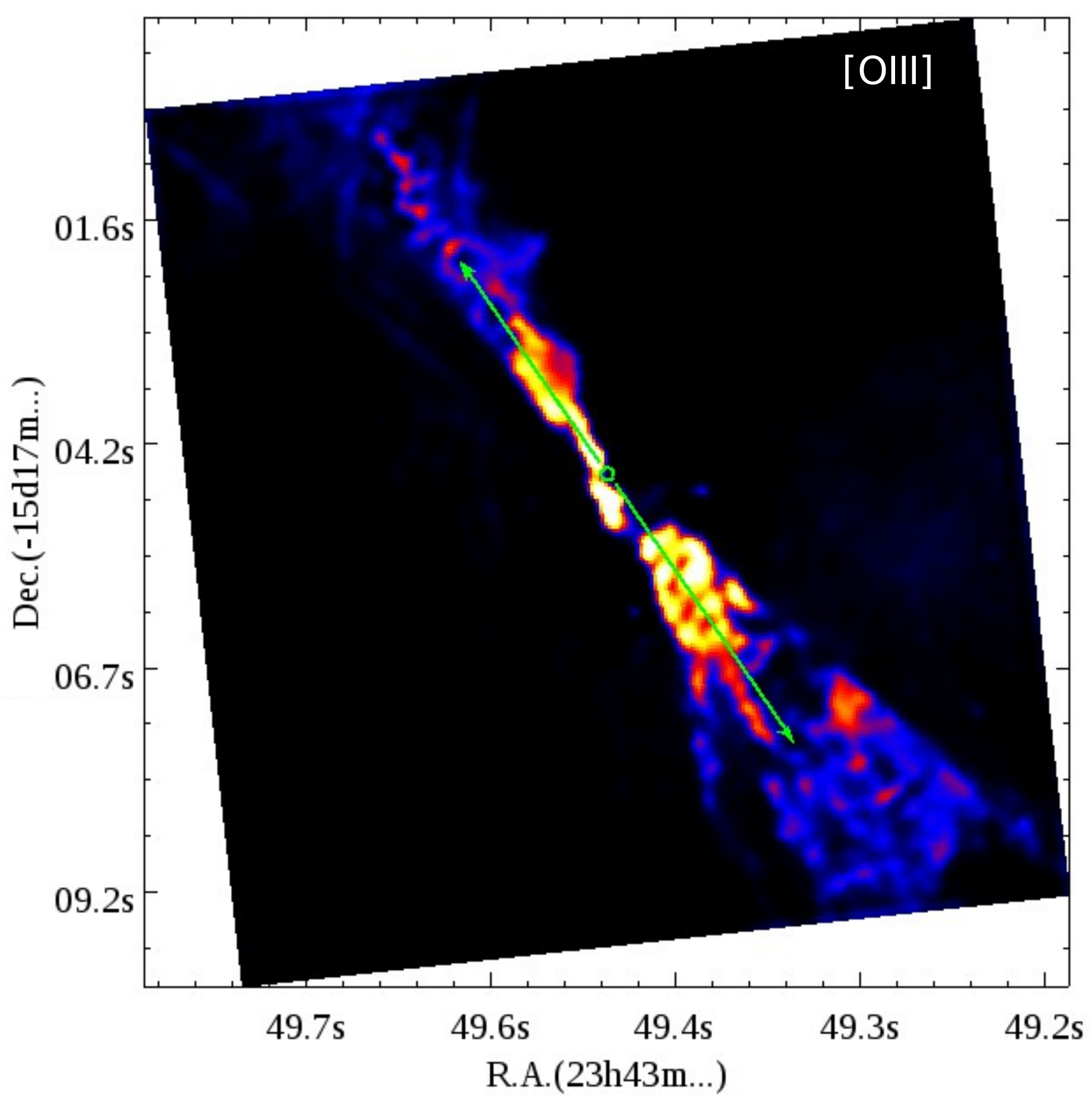}}
\end{minipage}
\begin{minipage}[c]{9.1cm}
\resizebox{\hsize}{!}{\includegraphics{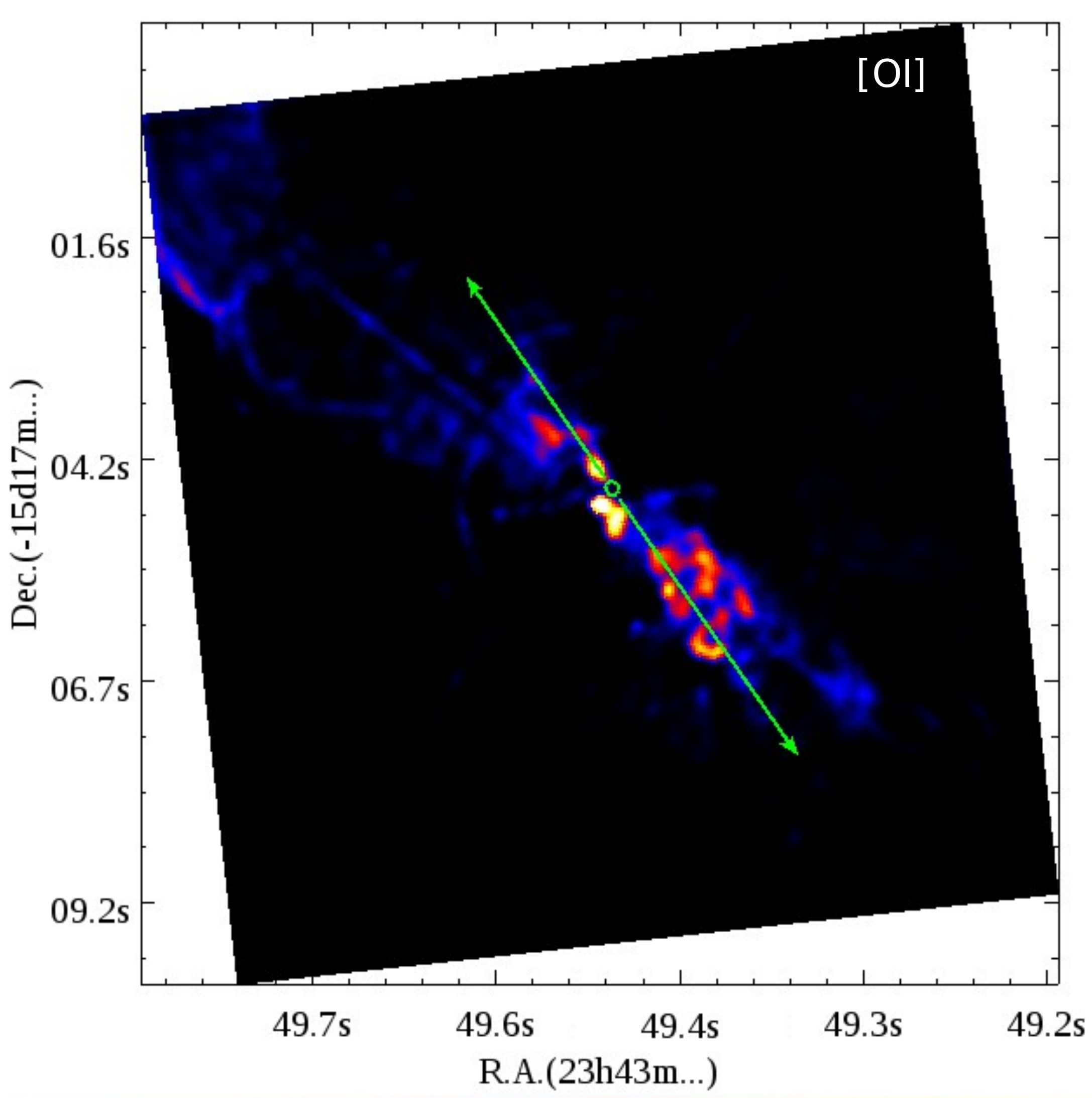}}
\end{minipage}

\begin{minipage}[c]{9.1cm}
\resizebox{\hsize}{!}{\includegraphics{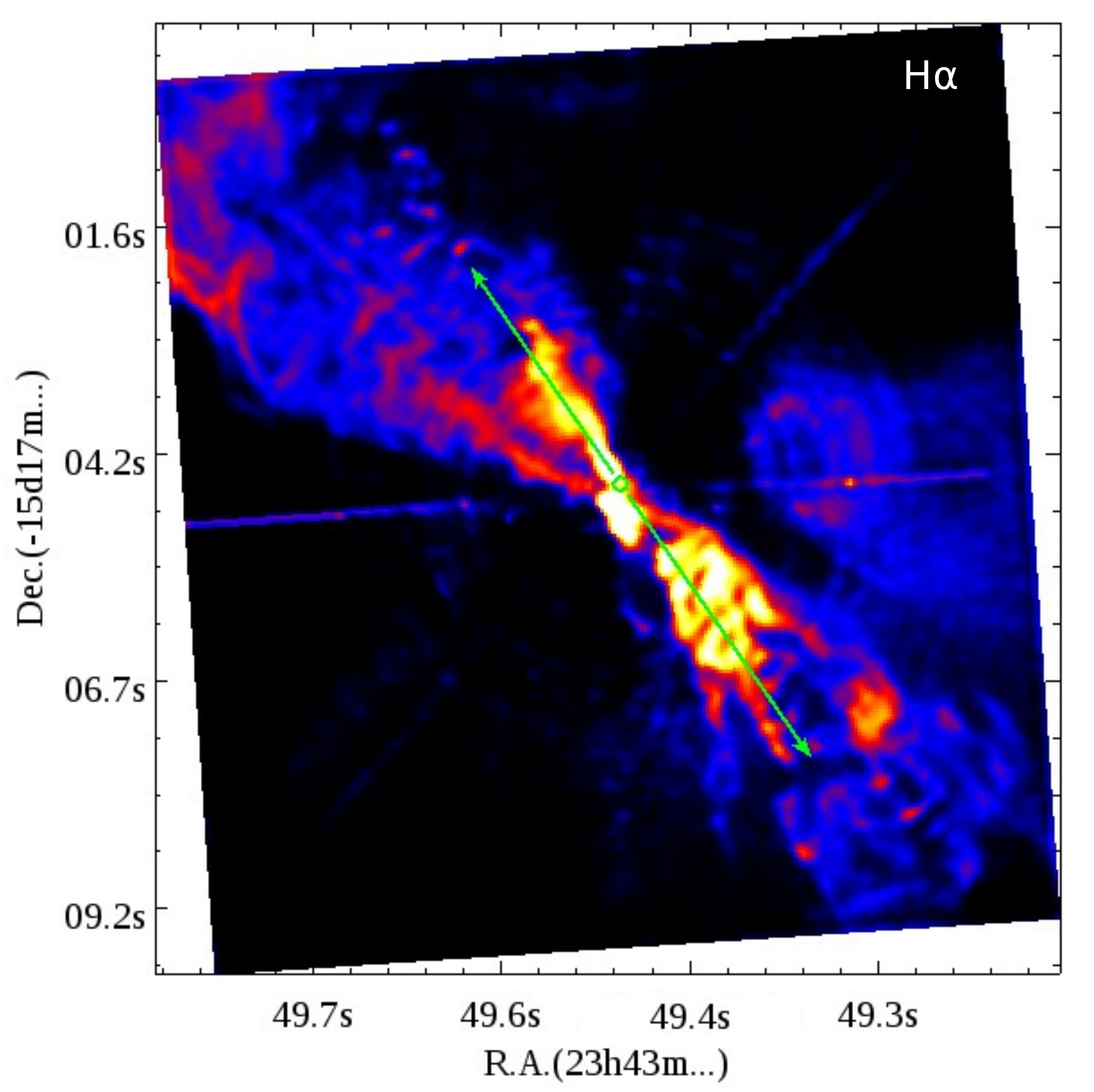}}
\end{minipage}
\begin{minipage}[c]{9.1cm}
\resizebox{\hsize}{!}{\includegraphics{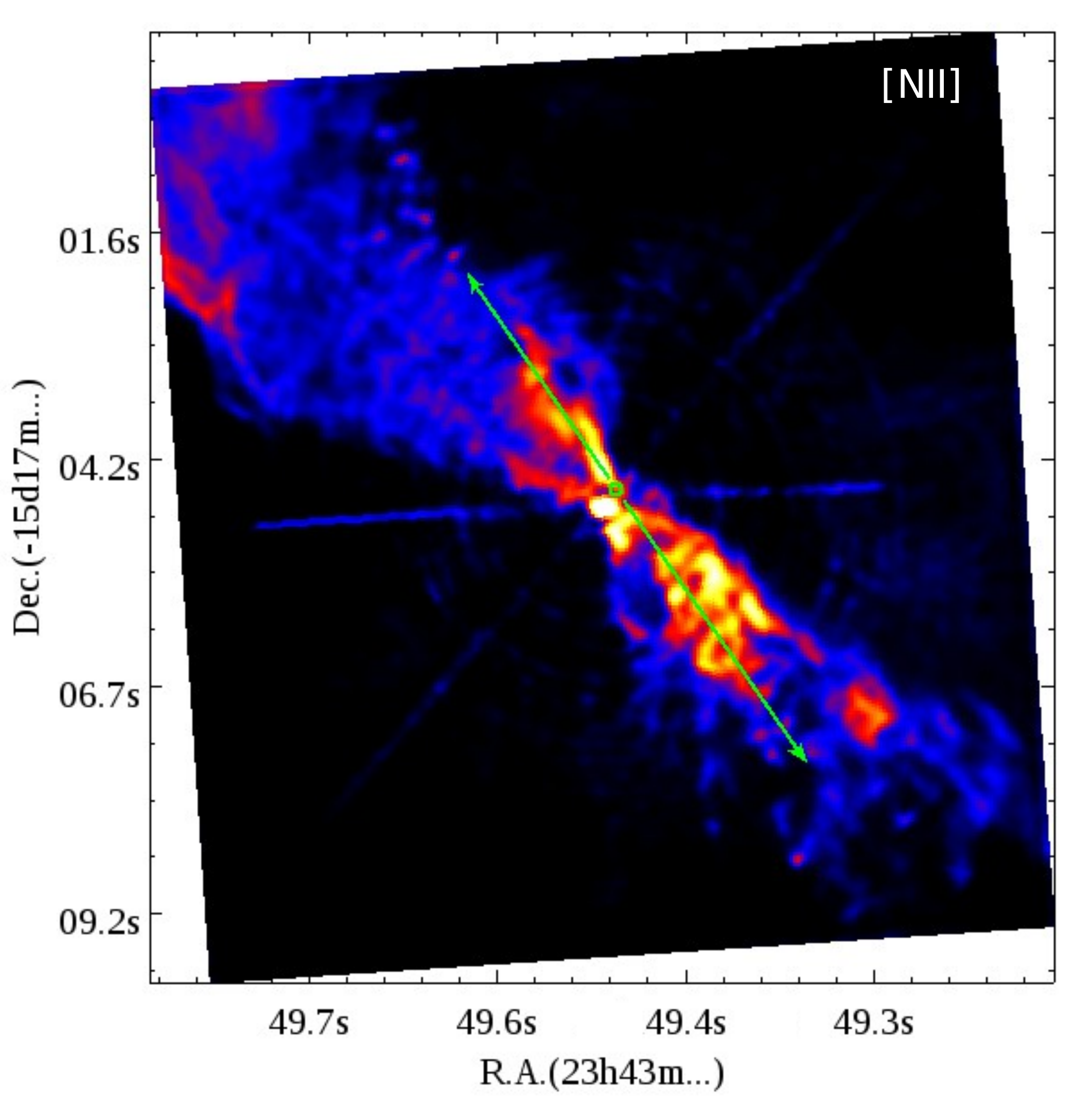}}
\end{minipage}

\begin{minipage}[c]{12cm}
\resizebox{\hsize}{!}{\includegraphics{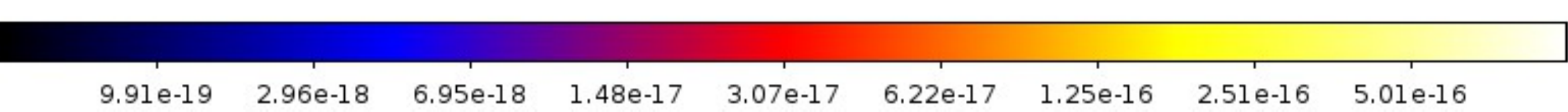}}
\end{minipage}
\caption{\label{flux_s} Narrow-band images of the central region with a size of $10\arcsec\times10\arcsec$ (filters given on the top left of each 
map). A PSF pattern has been subtracted. The colour scale for the fluxes ranges from 0 to $10^{-15}$ erg cm$^{-2}$ s$^{-1}$ \AA$^{-1}$. The adopted 
axis of the jet (green lines with arrows) was aligned according to a helical structure visible extended in NE direction on the [OIII]$\lambda$5007 
emission map; this structure probably represents a high-collimated jet. The jet base position (green circle) drawn from the adopted jet axis does not 
coincide with the position of the stellar components resolved in \citet{Sch17}.}
\end{figure*}

Reduced PSF-subtracted $10\arcsec\times10\arcsec$ images of the central region (Fig.~\ref{flux_s}) allow us to see a fine structure of the emission 
close to the jet source. Among the low-excitation lines, the [OI] emission is quite faint in the inner jet, whereas H$\alpha$ and [NII] are bright 
enough to display many morphological features here. At the same time, the H$\alpha$ map contains a roundish emission spot ($\sim 2-3$ 
arcsec to the west of the central source in Fig.~\ref{flux_l} and Fig.~\ref{flux_s}) which is not visible at other lines. Therefore, we assume that 
this is a kind of "image ghost". Also, it should be noted that the H$\alpha$ and [NII] maps may still contain some remaining artificial 
flux after the removal of the stellar PSF-pattern.

\begin{figure*}[t]
\centering
\begin{minipage}[c]{6.2cm}
\resizebox{\hsize}{!}{\includegraphics{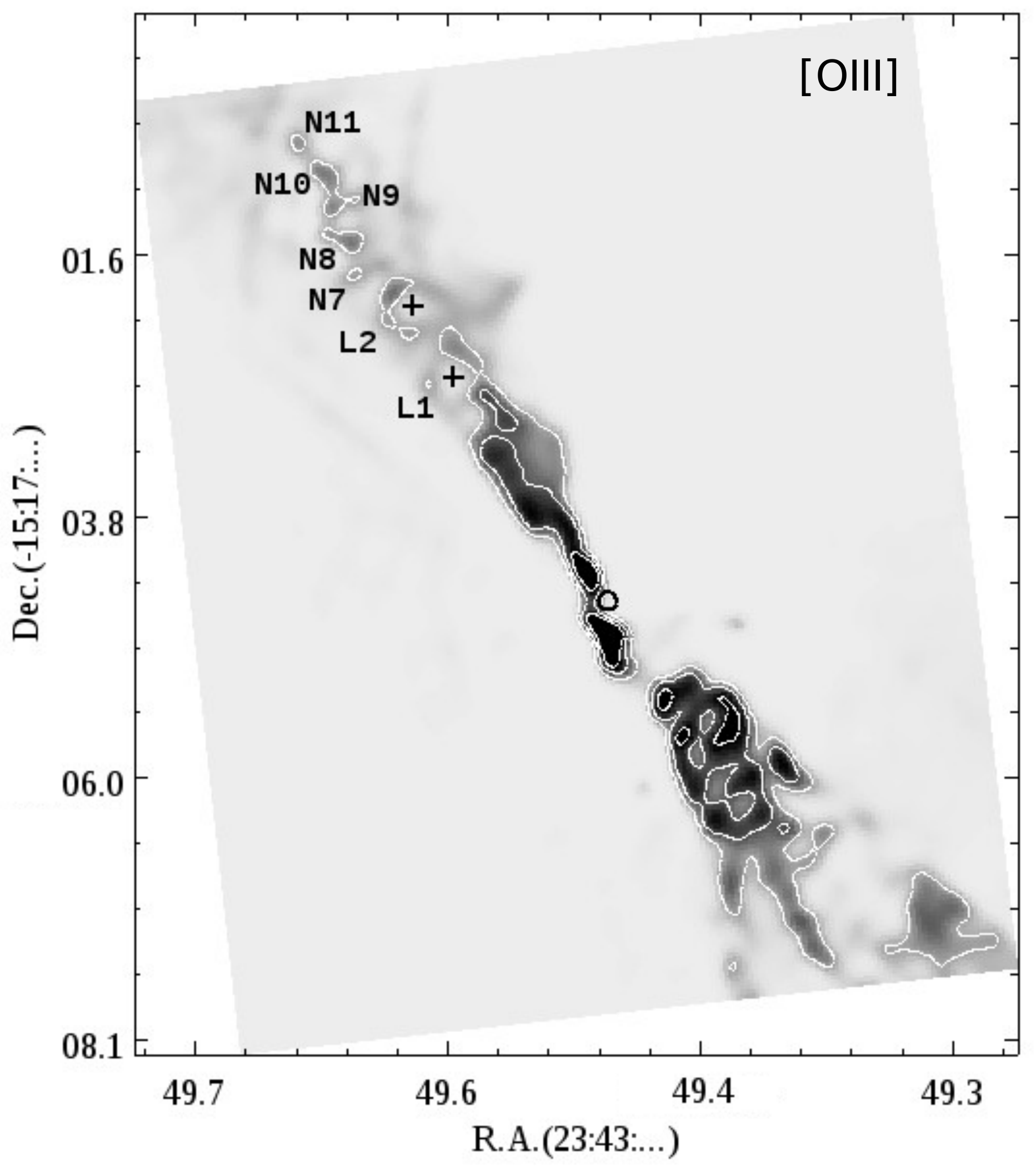}}
\end{minipage}
\begin{minipage}[c]{5.9cm}
\resizebox{\hsize}{!}{\includegraphics{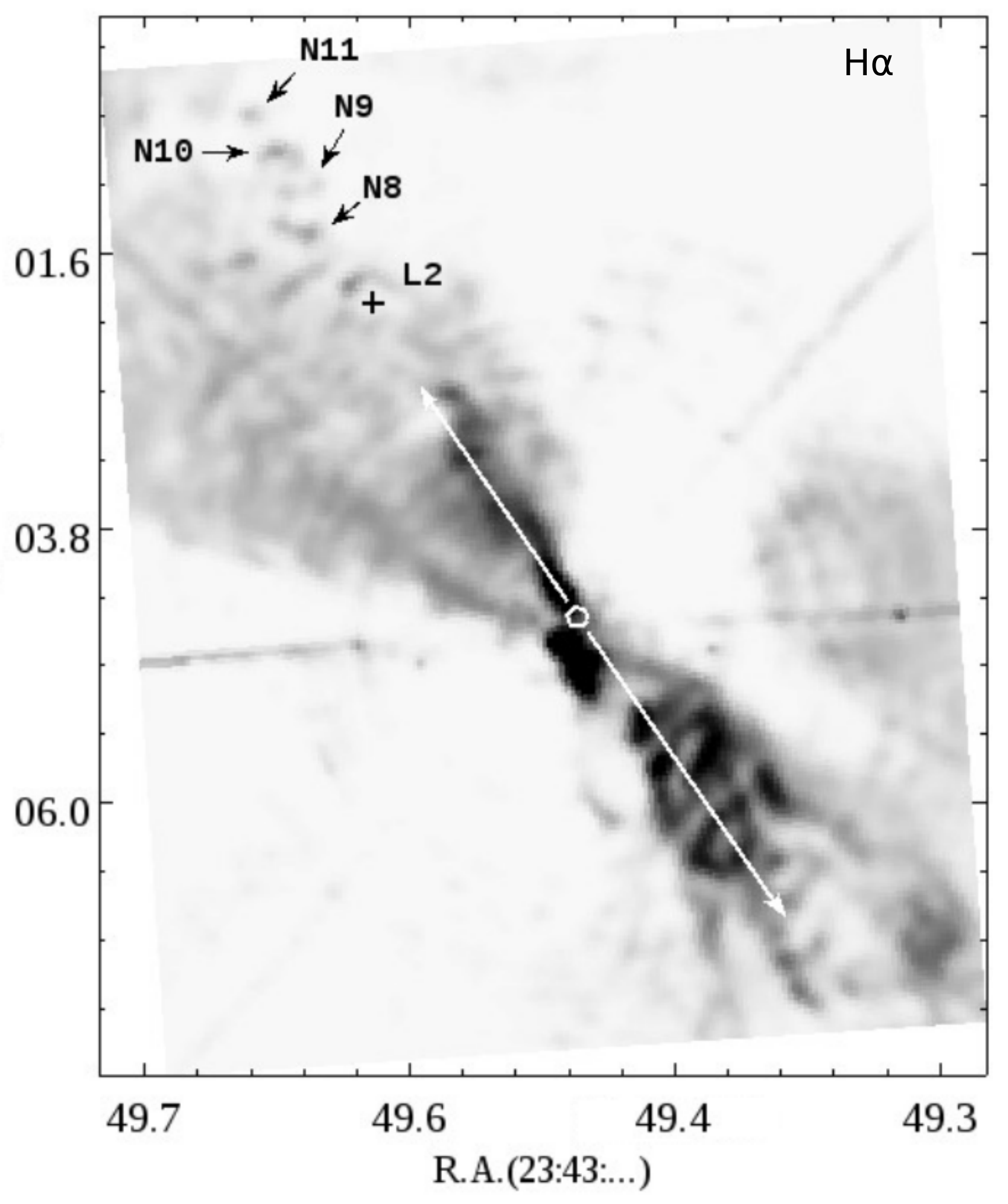}}
\end{minipage}
\begin{minipage}[c]{5.9cm}
\resizebox{\hsize}{!}{\includegraphics{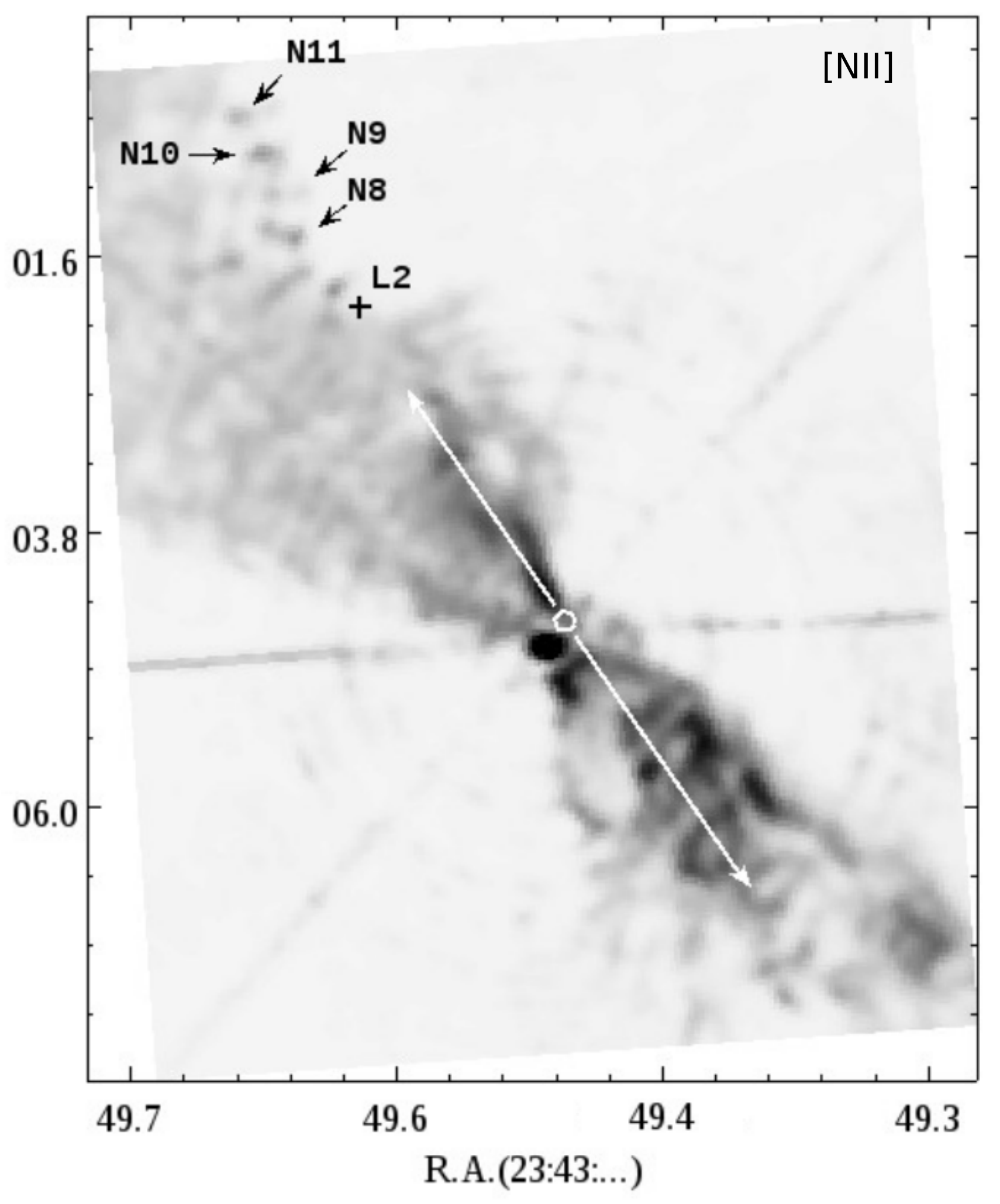}}
\end{minipage}
\caption{\label{injet} [OIII], H$\alpha$, and [NII] images of the central region demonstrating the inner, highly collimated jet; a PSF pattern was 
subtracted, but some faint PSF emission spikes  remain visible on the H$\alpha$ and [NII] images. The position of the adopted jet axis 
(white lines with arrows) is superimposed on the H$\alpha$ and [NII] images, as has been shown in Fig.~\ref{flux_s}. [OIII] contour levels are given 
for three levels from $10^{-17}$ to $10^{-15}$ erg cm$^{-2}$ s$^{-1}$ \AA$^{-1}$.  The [OIII] image exhibits the fine helical structure of the inner, 
highly collimated jet, which consists of a wiggling continuous flow, loops (L1 and L2), and the chain of small knots \citep[marked as N7--11 
following 
the numeration for "N"-knots from][]{Par94} This emission pattern can also be traced as loops on the H$\alpha$ and [NII] images. The [OIII] and 
H$\alpha$ images also show that whereas the starting position of the NE jet agrees well with the proposed location of R~Aqr stellar system, the 
starting position of the SW emission is shifted relative to the adopted jet axis by 0\farcs7.}
\end{figure*}

First of all, this high-excitation [OIII] line, as well as H$\alpha$ and [NII], reveal a highly collimated jet extending $\sim5\arcsec$ from the central 
source position (green circle) in the NE direction. Figure~\ref{injet} presents a close-up view of the inner jet on the [OIII], H$\alpha$, and [NII] maps. 
In the NE direction, the collimated jet consists of the bright wiggling outflow, several emission loops (L1-2), and a chain of small faint knots 
(N7-11) (see Fig.~\ref{injet} and also Fig.~\ref{flux_s}). These emission features are forming a helical pattern implying a precession of the 
jet or a propagation of emitting gas rotating around the jet axis. A "wiggling" outflow is often observed among young stellar jets and protostellar 
molecular outflows. We discuss this phenomenon in the Discussion Section. Unfortunately, in the SW direction we do not see any prominent chain of 
knots similar to that visible in the NE jet. At the same time, in the SW there is a wiggling spike which is roughly aligned with the proposed jet 
axis. Since the chain of NE knots in this highly collimated [OIII] structure in well aligned, we have adopted its position angle as the position of the 
jet axis (the green straight line in Fig.~\ref{flux_s}) and extended it towards the SW direction. However, this adopted axis does not intersect with 
the position of the center of the extracted PSF source, which should roughly coincide with the position of the binary system resolved by 
\citet{Sch17}. 

Beside the collimated jet in NE direction, the [NII] and H$\alpha$, as well as [OI], illustrate a wide opening angle flow of faint emission which is 
clearly  inclined with respect to the axis of the highly collimated jet, whereas we do not see such off-axis emission in [OIII].  While we can 
track the highly collimated component in [NII] and H$\alpha$ during the few first arcseconds, the wide-angle emission extends over about 15\arcsec, 
forming the main lobes visible on the large-scale maps (Fig.~\ref{flux_l}). The wide-angle [OI] emission is faint during the first several arcseconds, 
but it becomes brighter at the edge in the extended bending NE lobe. From the [NII] and H$\alpha$ images, we estimate the full opening angle 
of the flow at its starting base as $\sim25\degr$.

The SW outflow is less collimated than the NE counterpart and shows a larger opening angle. In the SW lobe, we can see several bright clouds in the 
[OIII] emission that \citet{Sch17} have described as bubbles or "zig-zag" pattern. We suggest that this pattern may also be interpreted as the 
"wiggling" motion of the emitting gas and its apparent shape can be formed due to an effect of projection or deformation of the gas flow distorted by 
some physical causes (see the Discussion Section). Unlike the NE jet, the SW outflow is clearly shifted to the west relative to the adopted jet axis 
and points to the position of the binary resolved by \citet{Sch17}. Since the same behaviour can be seen in all the observed lines, we conclude that 
this displacement of the gaseous flows is a real effect. Comparing the emission flows in different lines, we can see that the SW outflow in [OIII] 
shows the smaller opening angle than in the low-excitation lines.

\section{Emission line ratio maps}   
\label{rates}
Using the intensity maps presented above, we computed large-scale and small-scale line ratio maps for four line combinations --- [NII]/H$\alpha$, 
[OIII]/[OI], [NII]/[OI], and [OIII]/H$\alpha$. These line ratios are determined by several physical characteristics of the jet plasma. The [NII]/[OI] 
mostly depends on the hydrogen ionisation fraction and therefore it allows us to analyse the gas excitation along the outflow beams. The 
[OIII]/H$\alpha$ ratio is a sensitive probe of the outflow temperature and \citet{Sch17} used these lines to derive the gas temperature in the jet. 
Therefore, this map can be useful for the analysis of the detailed distribution of this parameter. The line ratios with H$\alpha$ are convenient 
because the Balmer line is usually bright in the spectra of stellar jets. Thus, the comparison of the H$\alpha$ line, which is produced by different 
mechanisms, with the forbidden lines, which are excited by a single mechanism, allows us to analyse the distribution of gas excitation produced by 
different processes and to study which mechanism dominates. The determination of the outflow gas parameters and their analysis will be presented 
in a following paper. Thus, in the current paper we only discuss the distribution of some line ratios.
\begin{figure*}
\centering
\begin{minipage}[c]{9.1cm}
\resizebox{\hsize}{!}{\includegraphics{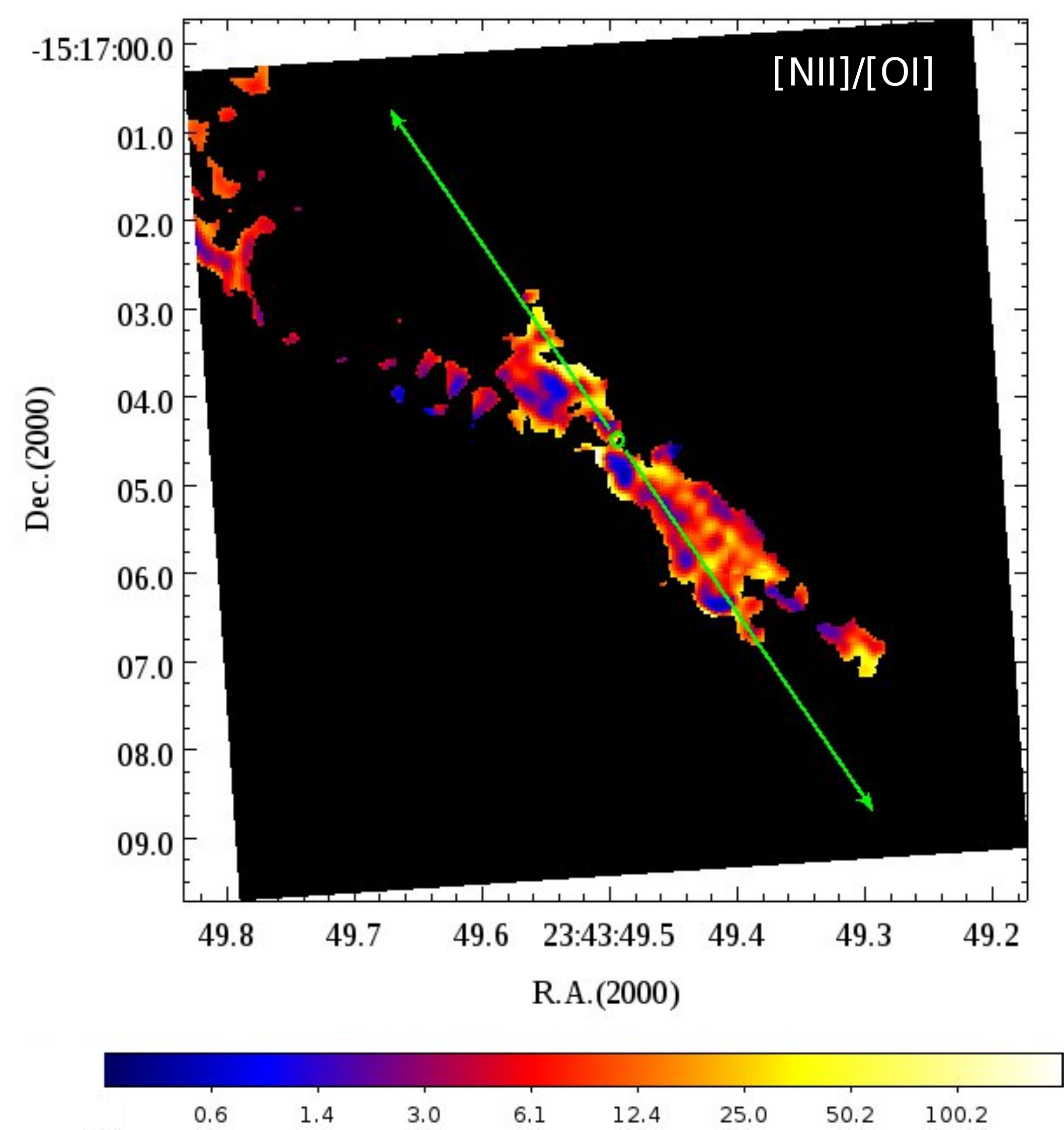}}
\end{minipage}
\begin{minipage}[c]{9.1cm}
\resizebox{\hsize}{!}{\includegraphics{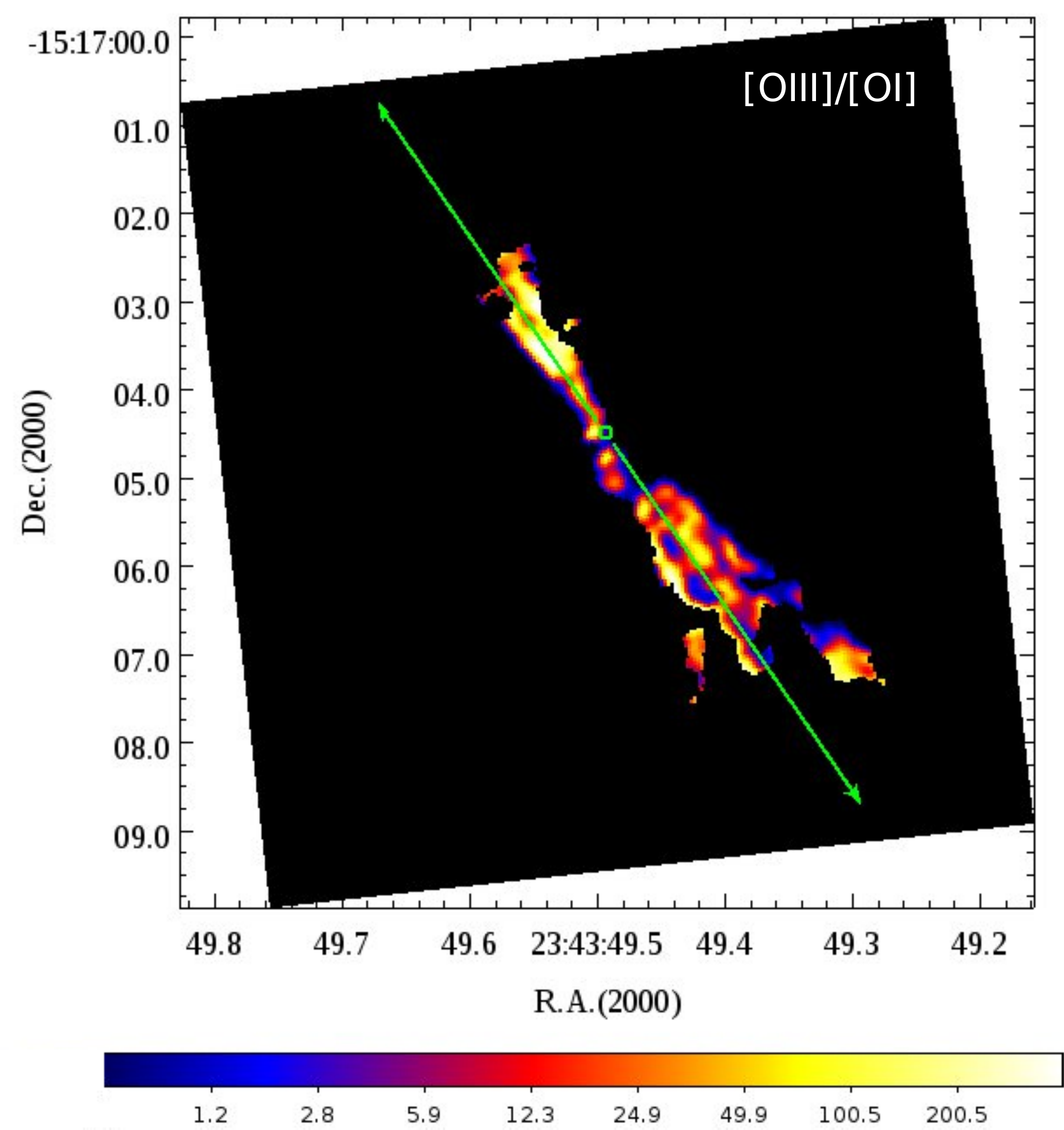}}
\end{minipage}

\begin{minipage}[c]{9.1cm}
\resizebox{\hsize}{!}{\includegraphics{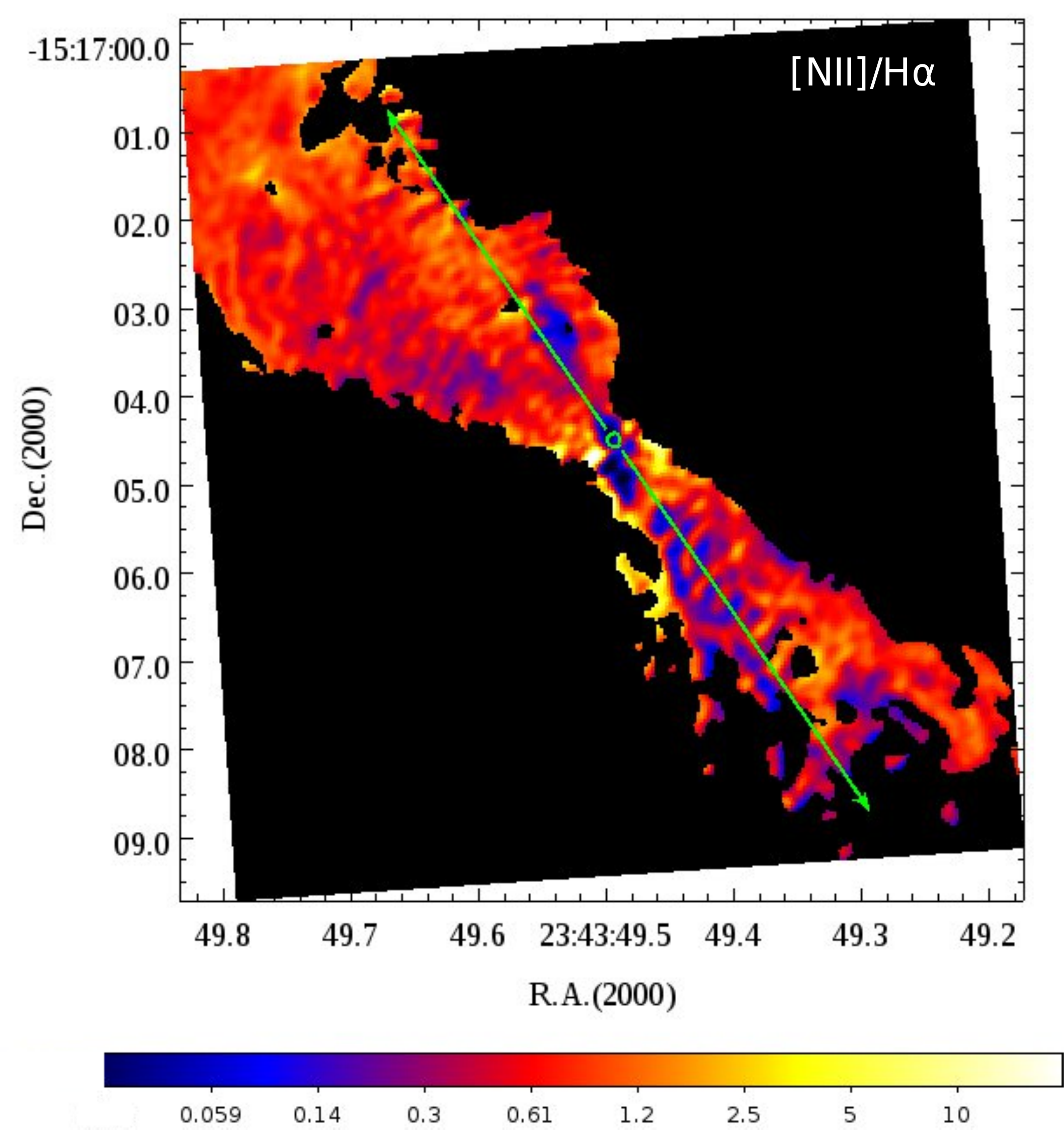}}
\end{minipage}
\begin{minipage}[c]{9.1cm}
\resizebox{\hsize}{!}{\includegraphics{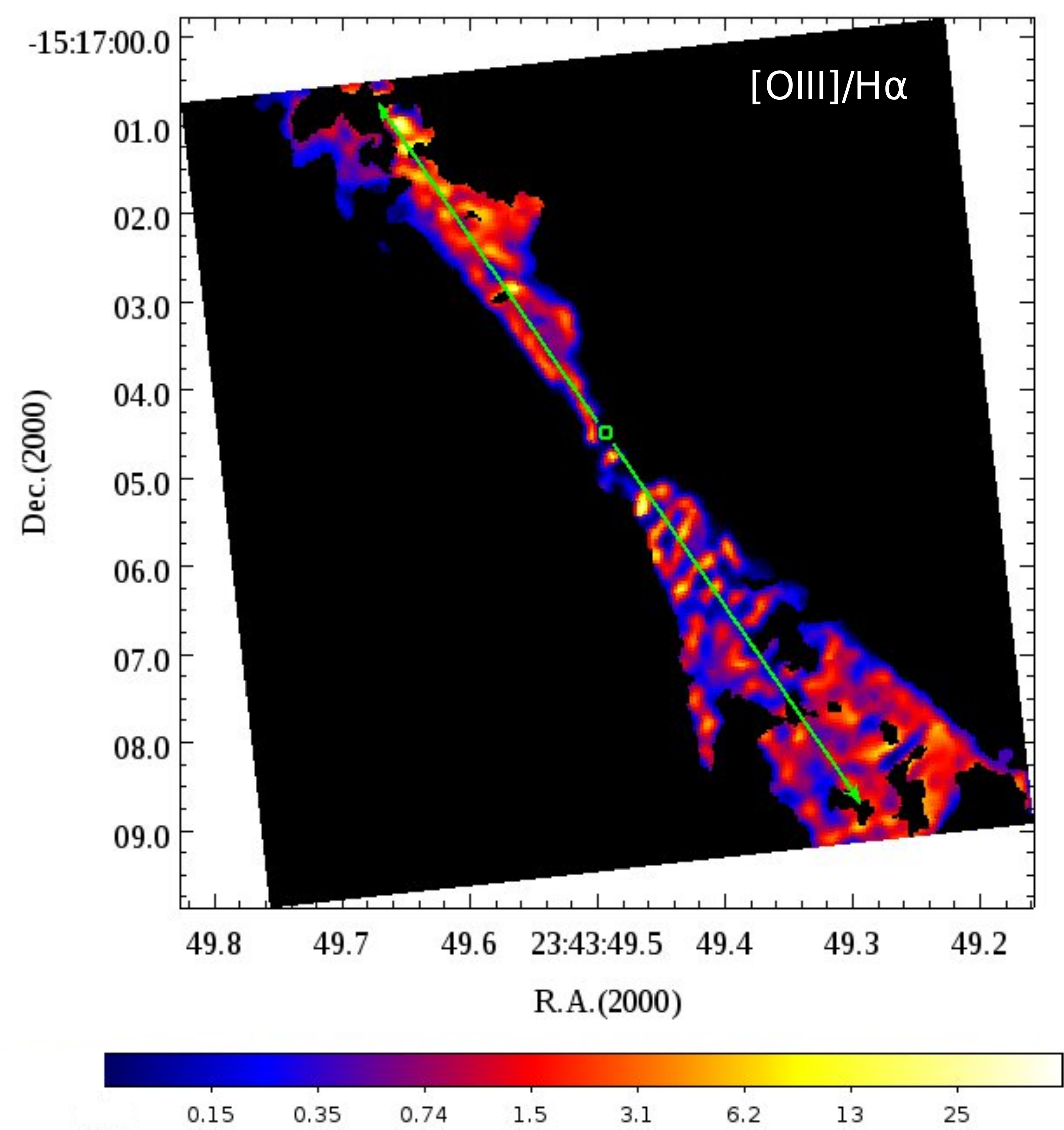}}
\end{minipage}
\caption{\label{ratio_s} Line ratio maps of the $10\arcsec\times10\arcsec$  central region (the lines given on the top left of each map). The 
position of the adopted axis of the highly collimated jet (green lines with arrows) is superimposed following Fig.~\ref{flux_s}.} \end{figure*}

For computing these maps we required a per-pixel S/N of 3 to avoid deriving less meaningful ratios from overly faint fluxes. The position of the 
inner jet axis is the same as adopted from intensity maps (Fig.~\ref{flux_s}). The $10\arcsec\times10\arcsec$ line ratio maps around the inner jet are 
presented in Fig.~\ref{ratio_s}. The adopted position of the axis of the highly collimated jet (green lines with arrows) is following that from the 
[OIII] flux map (Fig.~\ref{flux_s}).

1. [NII]/[OI]: If we compare averaged values, the ratios in both lobes are similar and around 10. If we consider the ratio values along the jet axis 
direction, however, the ratio within the first two arcseconds along the NE lobe is much stronger ($\sim50$) than it is in the SW lobe, where the ratio 
is about one order lower, $\sim5$. At the same time, the ratio for the reliable fluxes can be traced over longer distances in the SW lobe than in the 
NE direction. In the NE lobe we were able to derive the line ratios within $\sim1\arcsec$, whereas in the opposite direction they are traced up to 
$2\farcs5$. This asymmetry may reflect a difference in the ionisation rate, which is often observed in the YSO jets. On smaller scales, one can see 
that the emission field exhibits fast variations of balance between the [NII] and [OI] fluxes. For example, at the outer edge of the jet beams one can 
see bright rims with the higher values of the ratios.

2. [OIII]/[OI]: This ratio map of the oxygen lines with different excitation level shows that these lines probably originate from slightly different 
regions. A high ratio region is aligned along the axis of the collimated jet whereas there are long rims with a low ratio on opposite edges of both 
jet lobes --- on the eastern edge for the NE jet and on the western edge for the SW jet. A comparison of the spatial location of the [OIII]- and 
[OI]-lobes clearly shows a displacement of areas where this emission originates. Moreover, the flux maps clearly show the existence of off-axis [OI] 
emission of the flow which is also observed in other low-excitation lines, while [OIII] reveals only the axisymmetric emission. Therefore, the 
[OIII]/[OI] map mostly represents the values around the collimated jet axis. 
\begin{figure*}
\centering
\begin{minipage}[c]{9.2cm}
\resizebox{\hsize}{!}{\includegraphics{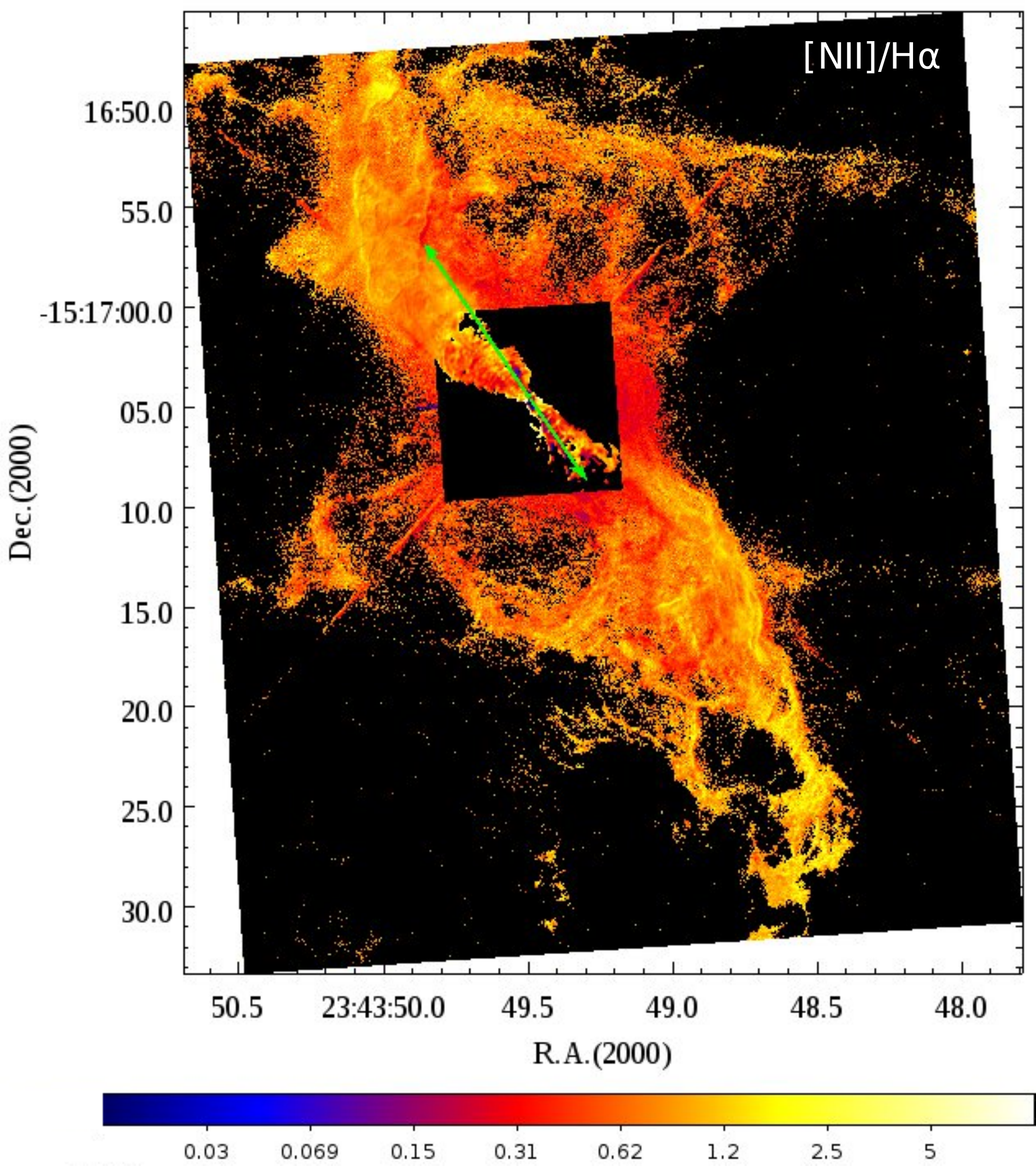}}
\end{minipage}
\begin{minipage}[c]{8.5cm}
\resizebox{\hsize}{!}{\includegraphics{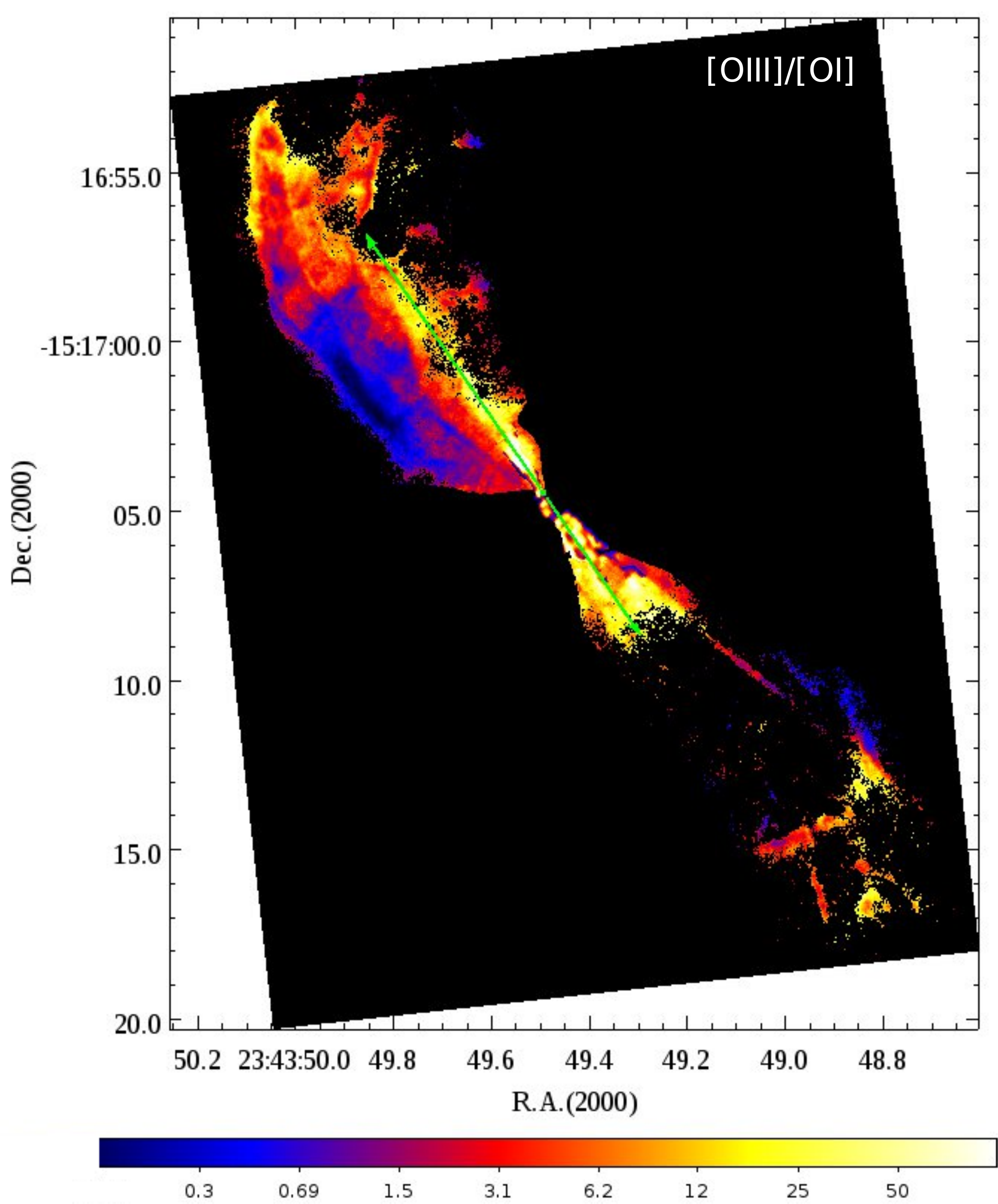}}
\end{minipage}

\begin{minipage}[c]{8.6cm}
\resizebox{\hsize}{!}{\includegraphics{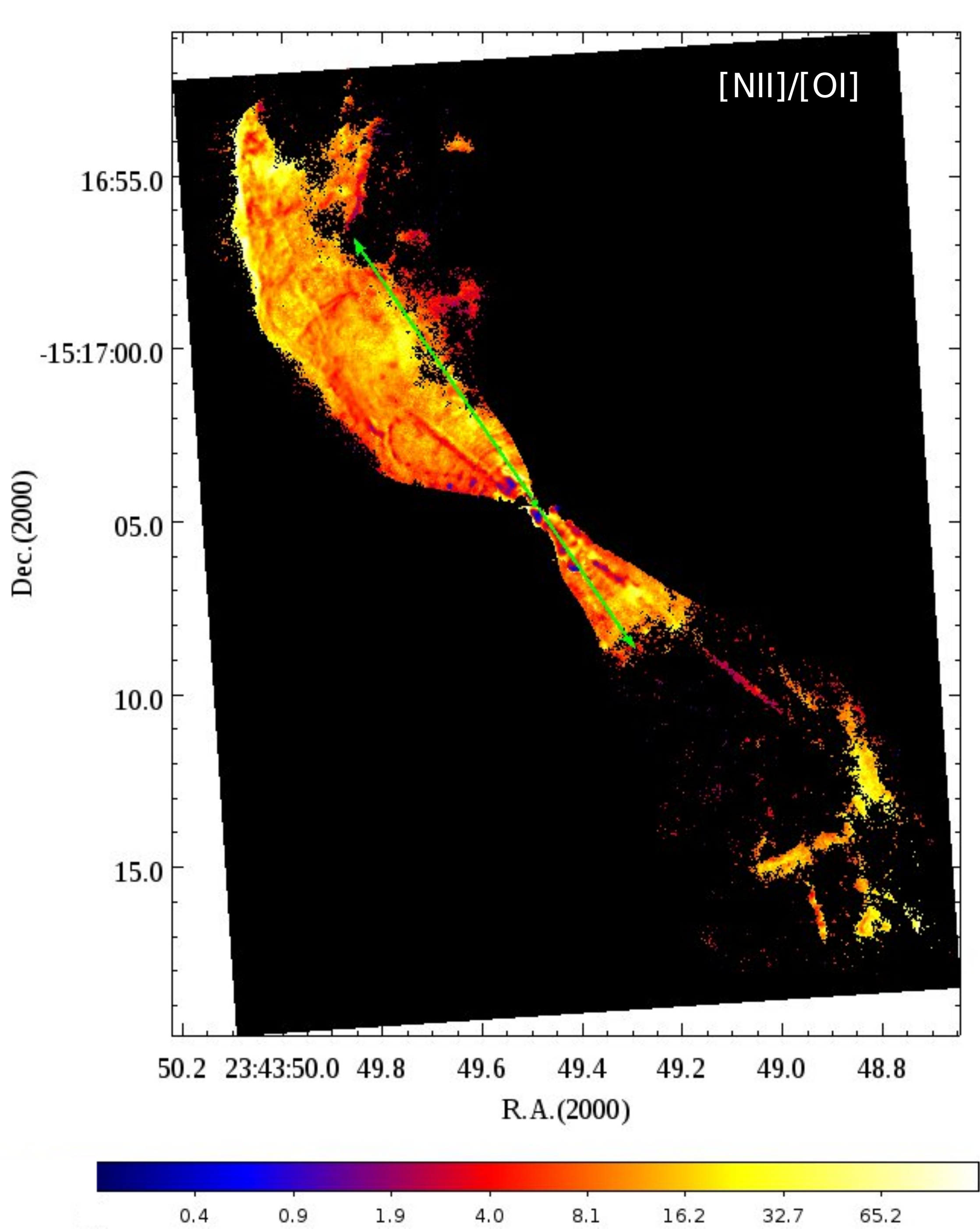}}
\end{minipage}
\begin{minipage}[c]{9.0cm}
\resizebox{\hsize}{!}{\includegraphics{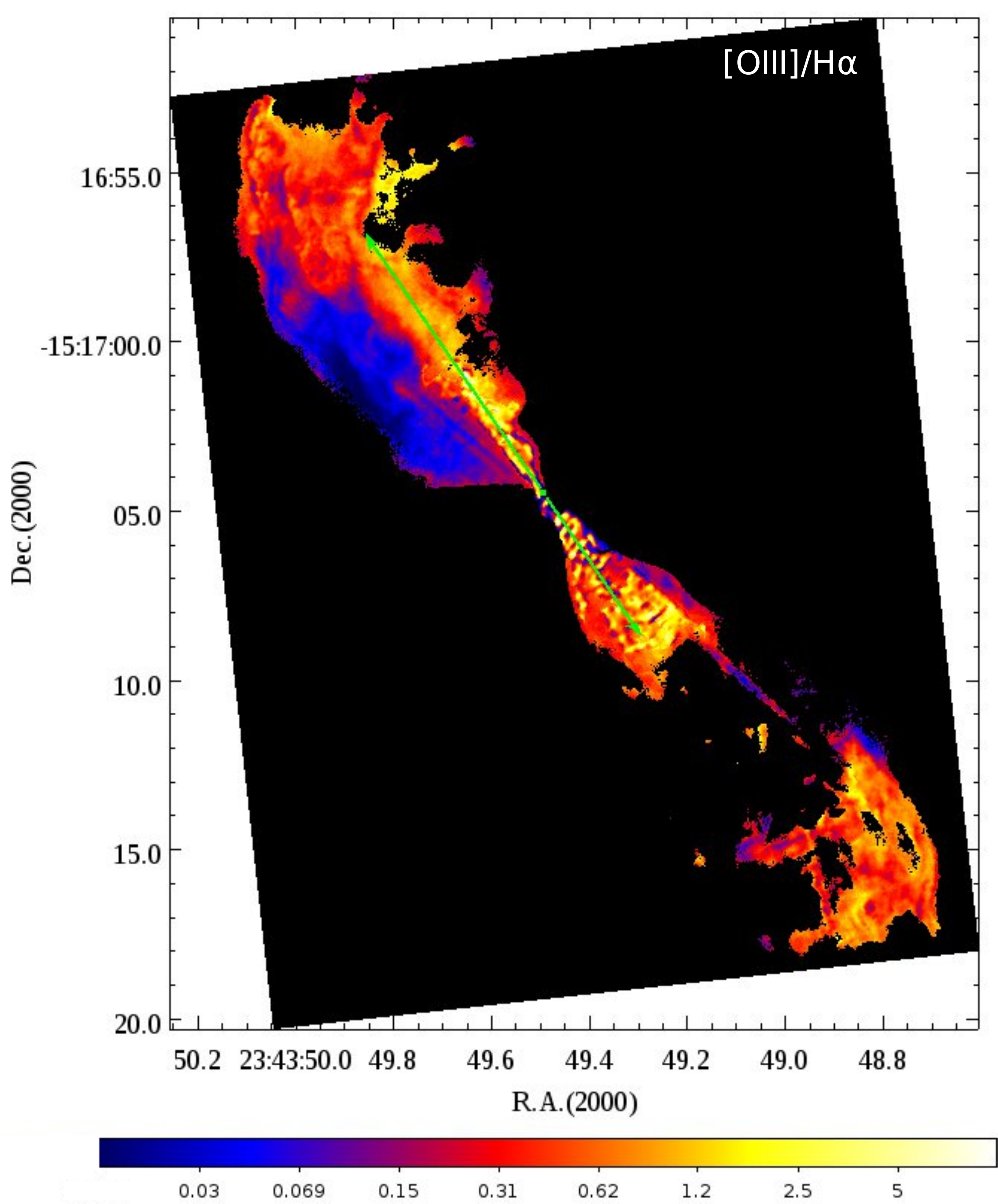}}
\end{minipage}
\caption{\label{ratio_l} Line ratio $\sim30\arcsec\times20\arcsec$ maps of two filters (the lines given on the top left of each map). The position 
of the adopted axis of the highly collimated jet (green lines with arrows) is superimposed according to Fig.~\ref{flux_s}.}
\end{figure*}

3. [NII]/H$\alpha$: This ratio map based on H$\alpha$ (as well as [OIII]/H$\alpha$) shows in general lower absolute values in comparison to the 
previously discussed line ratio maps. This map shows that the off-axis emission both in [NII] and in H$\alpha$ comes from similar areas in the NE lobe 
and the off-axis emission can also be traced in the SW part with similar ratio values. Along the collimated jet the [NII]/H$\alpha$ ratio is in 
general lower than in the off-axis areas. In particular, this ratio is low in the SW lobe, where it delineates the "zig-zag" pattern. In the upper 
part of the NE collimated jet (N8 -- N11 in Fig.~\ref{injet}) the ratio exhibits a clumpy pattern that consists of a chain of emission knots and 
loops.

4. [OIII]/H$\alpha$: Due to the high collimation of [OIII], the distribution of [OIII]/H$\alpha$ can only be studied in the region around the jet 
axis. The ratio of these bright emission lines shows a fine structure which follows the complex emission field structure and exhibits three zones --- 
the low (blue zones), the moderate (red), and the high ratio values (yellow). This structure indicates the different areas where the lines originate 
from. Deciphering why the opening angles in the opposite directions are different requires an additional analysis. At the same time, the 
distribution of the ratio values looks more homogeneous along the NE jet than in the opposite direction.

The computed $30\arcsec\times20\arcsec$ line ratio maps (Fig.~\ref{ratio_l}) represent the mosaic of large-scale fields superimposed with the 
PSF-subtracted $10\arcsec\times10\arcsec$ central areas.

1. [NII]/H$\alpha$: Both the H$\alpha$ and [NII] emission images contain very extended areas of faint emission which were unfortunately removed from the $10\arcsec\times10\arcsec$ ratio maps in the process of the PSF-subtraction. Therefore, the superposition of the large- and small-scale fields has led to wiping out the line ratio value in the central region of the ratio map. Nevertheless, we decided to keep the superimposed map to show how the flux ratio transforms from the inner collimated jet to the peripheral outflow areas. The [NII]/H$\alpha$ ratio reveals the richest morphology of the emission structures. Its distribution shows that the zones of [NII] and H$\alpha$ excitation coincide very well and many dynamical emission structures can be traced by following these lines. This ratio map outlines various gaseous features such as long arcs and large loops which reveal the complex behaviour of this bipolar outflow. The absolute values of [NII]/H$\alpha$ are quite low, ranging mostly from 0.3 to 2, and the ratio has a trend of growing with distance from the central jet source. The highest values of this ratio can be seen at the emission arcs on the outer edges of the wide-angle outflow, especially in the SW lobe. Mean ratio values are similar for both the 
inner and outer outflow.

2. [OIII]/[OI]: This map represents the ratio of forbidden oxygen lines with different ionisation levels --- the low-excitation [OI] and the higher-excited 
[OIII]. This ratio displays relatively high values around the inner jet axis (yellow areas), seen in particular in the NE lobe. At the same time, the ratio values gradually decrease with distance from the central source. At the beginning of the NE jet the ratio values are about 
70--90, while at the distance of 5--6$\arcsec$ from the centre, they drop down to 10--15, with average values of about 30--40. In the SW direction, such a gradient is also observable, although it is not so prominent. In addition, the ratio drops to lower values with growing distance from the axis of the collimated jet. This effect is most prominent in the external parts of the NE lobe (blue area) and visible also in the SW part as a red rim along the bright part of the inner jet and as several faint blue spots along the distant arc located at 10$\arcsec$. 

3. [NII]/[OI]: The bulk of the gas that is sufficiently strong to allow us to measure a reliable line ratio is clearly located off the inner jet axis. In the NE jet the emitting gas is shifted towards the NE direction, whereas in the opposite lobe it is shifted towards the SW. This same effect is also observable for the [OIII]/[OI] and [OIII]/H$\alpha$ ratios. The wide-angle NE outflow shows very fine excitation structure, with the lower values of the ratio forming a mesh of small-scale arcs and ridges over the more homogeneous field with the higher ratio values. The most distant outer edge of the lobe has the  shape of a long bright rim. On the other hand, the [OI] in the SW jet is bright for only the first few arcseconds and beyond this distance we see only a long arc located at 10$\arcsec$ from the jet source. The [NII]/[OI] shows quite a narrow range of ratio values, mostly between 2 and 15.

4. [OIII]/H$\alpha$: This ratio exhibits a behaviour similar to [OIII]/[OI]. We notice the prominent jet-like collimated area of high values aligned with 
the inner jet axis in the NE direction, where it can be traced up to 8$\arcsec$ from the centre. Both jet lobes also reveal the zones where the ratio 
values are lower than around the adopted inner jet axis. On the opposite lobes these zones are located at the opposite edges of the bipolar outflow. The range 
of this ratio is quite narrow and the absolute values are, in general, low --- around unity along the jet axis and 0.2--0.5 outside this region.

Summarising this review of the flux emission and line ratio maps, one notices that the morphology of high-excitation ([OIII]) and low-excitation ([OI], [NII], H$\alpha$) emission  is different. The [OIII] emission is extended mostly along the axis of the inner collimated jet and it is more highly collimated than the low-excitation emission. On the other hand, the [OI], [NII], and H$\alpha$ emission also concentrates along this axis only at the beginning of the jet, within $\sim2\arcsec$ in both directions. Further out, the low-excitation emission forms the curved lobes with the bend pointing in opposite directions. The bulk of the emission in the lobes is also displaced from the inner jet axis in the same directions.

\section{Proper motions} 
\label{PM}
The R~Aqr jet has now been observed for many years. Thus it was possible to detect the proper motion (PM) of its gaseous knots and to measure it with 
the help of high-resolution observations. This way, \citet{Leh92} used radio observations with the VLA to find the gas proper motion, while 
\citet{Hol93} measured the PM of the features in the R~Aqr flow from the early HST imagery obtained around 1990. In the latest study of the 
high-resolution images, \citet{Sch17} noticed an inferred outward motion of gas clouds from images separated by only two months and estimated a motion 
on the order of 40 mas yr$^{-1}$ within $0\farcs7$ from the jet source. For a new, more detailed PM study on a larger time baseline, we combined the 
2013/14 HST observations with the available archival optical HST data of the R~Aqr jet. Since the high-resolution HST data exhibit a wealth of new 
features, we intend to publish this analysis in full detail in a separate paper, but here we summarise our basic conclusions on the PMs.

\begin{figure}
\centering
\resizebox{\hsize}{!}{\includegraphics{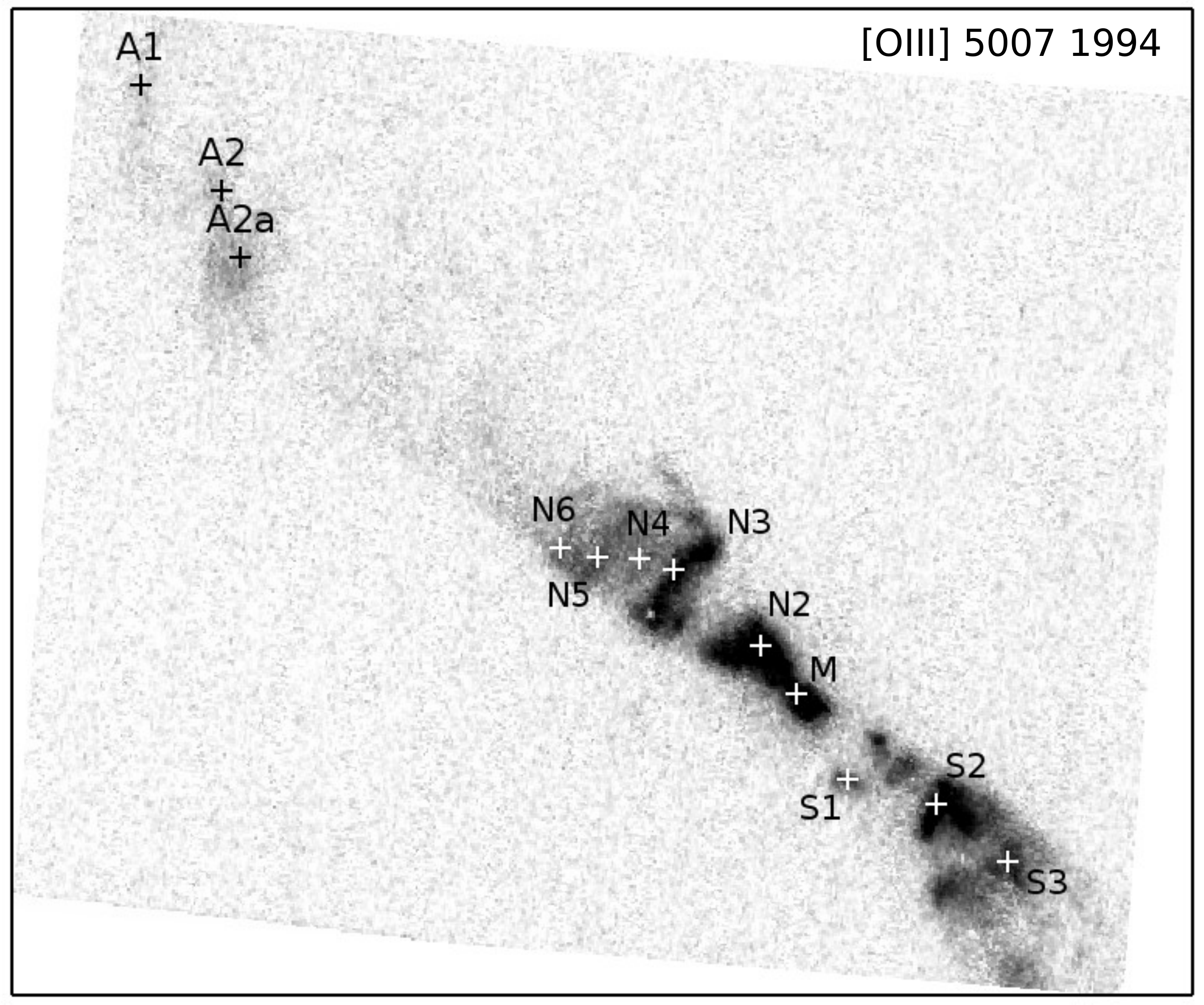}}
\caption{\label{knots_arx} Example of the archival [OIII]$\lambda$5007 image (F501N) of the central region of R~Aqr obtained in 1994. The 
identified knots are labelled following \citet{Par94}. After inspection of all archival images, we have identified another knot marked on the image 
as A2a.}
\end{figure}

Between 1990 and 1998 R~Aqr was observed a number of times with HST and its FOC, HRS, and NICMOS instruments. Therefore, these long-term observations 
spanning 24 years of epoch difference can, in principle, allow us to measure the PM of the knots with very high accuracy. A detailed comparison of the 
2013/14 HST observations with the 1990--98 data, however, shows that most of the jet knots near the core of the R~Aqr system have considerably evolved 
since 1990 and their structure has changed significantly. Moreover, this rapid evolution can be seen even on short time scales. On such high-resolution images 
separated by only two months, \citet{Sch17} found that some H$\alpha$ knots already showed noticeable brightness changes. Similarly fast evolution of 
the knot brightness is also detected in X-rays \citep{Kel07}. In addition to that, the WFC3/UVIS camera has a much higher sensitivity compared to the 
older HST cameras leading to the detection of fainter emission details. Due to these massive changes in the visible emission structures, we could only 
identify the prominent knot N3 as denoted by \citet{Par94} in the 2013/14 HST data. Consequently, we could derive the PMs for most of the knots only 
from the archival 1990--94 HST data.
\begin{table}
\centering
\caption{\label{PM_tab} Proper motion of the emission knots from 1990--2014 HST observations.}
\begin{tabular}{@{\hspace{3mm}}r@{\hspace{3mm}}|l@{\hspace{3mm}}l|@{\hspace{3mm}}c@{\hspace{3mm}}l}
\hline\hline
Knot & r$_{jet}$ & PA  & PM$_{rel}\pm$rms  & JD$_0\pm$rms \\
     & \multicolumn{2}{c|@{\hspace{3mm}}}{(epoch=1994.8)}        &    &              \\
     & [\arcsec] & [\degr] & [\arcsec yr$^{-1}$] &  \\ 
\hline 
\multicolumn{5}{l}{NE jet}\\
 N2   &  0.27 &       &  $-0.018\pm0.004$ &                      \\
 N3   &  0.91 & 45.5  &  $ 0.037\pm0.001$ &  $ 2440329\pm  344 $ \\
 N4   &  1.10 & 51.8  &  $ 0.013\pm0.007$ &  $ 2418146\pm23809 $ \\
 N5   &  1.32 & 55.6  &  $ 0.039\pm0.007$ &  $ 2437410\pm 2066 $ \\
 N6   &  1.52 & 58.2  &  $ 0.040\pm0.009$ &  $ 2435895\pm 3149 $ \\
 A1   &  4.82 & 46.8  &  $ 0.174\pm0.015$ &  $ 2439572\pm  846 $ \\
 A2   &  4.16 & 49.0  &  $ 0.125\pm0.026$ &  $ 2437633\pm 2451 $ \\
A2a   &  3.85 & 49.4  &                   &                      \\ 
\hline
\multicolumn{5}{l}{SW jet}\\
 S1   &  0.58 & 213.9 &  $ 0.086\pm0.008$ &  $ 2447199\pm 181 $ \\
 S2   &  0.98 & 230.4 &  $-0.007\pm0.008$ &                     \\
 S3   &  1.54 & 230.9 &  $ 0.055\pm0.011$ &  $ 2439491\pm1932 $ \\
\hline
\end{tabular}
\end{table}

To extract accurate astrometry from the archive images, we subtracted an instrumental point spread function (PSF) pattern of these images and reduced 
their world coordinate system (WCS) to a single epoch. In Fig.~\ref{knots_arx} we present the knots identified on an archival [OIII]$\lambda$5007 
image obtained in 1994. The designation of these knots is mostly following \citet{Par94}. An additional emission feature located near the A2 knot and 
seen on several archival images is labelled as A2a here. Table~\ref{PM_tab} presents the knot positions derived for the 
1994 epoch when the knots are firmly identified (Fig.~\ref{knots_arx}). All the positions are given in polar coordinates --- angular distance 
from the jet source (r$_{jet}$) and position angle (PA). Of these knots, only the position of N3 has also been derived from the 2013/14 HST 
images --- r$_{jet}=1\farcs67$ and PA=46\fdg9, which show a considerable angular propagation of this knot in comparison with 1994. We also measured 
the positions of the knots for all the available epochs, and calculated their PM indicating the motion relative to the jet source 
(Table~\ref{PM_tab}). The analysis of the motion of the emission features shows that, in general, the knot velocities increase with increasing 
distance from the jet source, which may imply the acceleration of the jet material with distance. An alternative explanation could be that the 
ejection velocity was higher in past, but has been decreasing with time. The change of the PMs agrees with the result of \citet{Mak04} who also found 
increasing PM of the knots by using radio observations with the VLA taken in 1992--1999. \citet{Mak04} concluded that within the first 1\arcsec\,from 
the jet base the jet consists of slowly moving gas, but at a distance $\sim1\arcsec$ from the jet origin the first shock is created, the gas is 
accelerating after the shock front and then  travels on a ballistic orbit. The average of our PM values of the knots is about 0\farcs07 per year, 
which corresponds to a spatial velocity of about 80 km/s for the distance and inclination of the jet. This result shows that the velocity of the gas 
in this outflow is generally similar to that reported in young stellar jets.

\begin{figure}
\centering
\resizebox{\hsize}{!}{\includegraphics{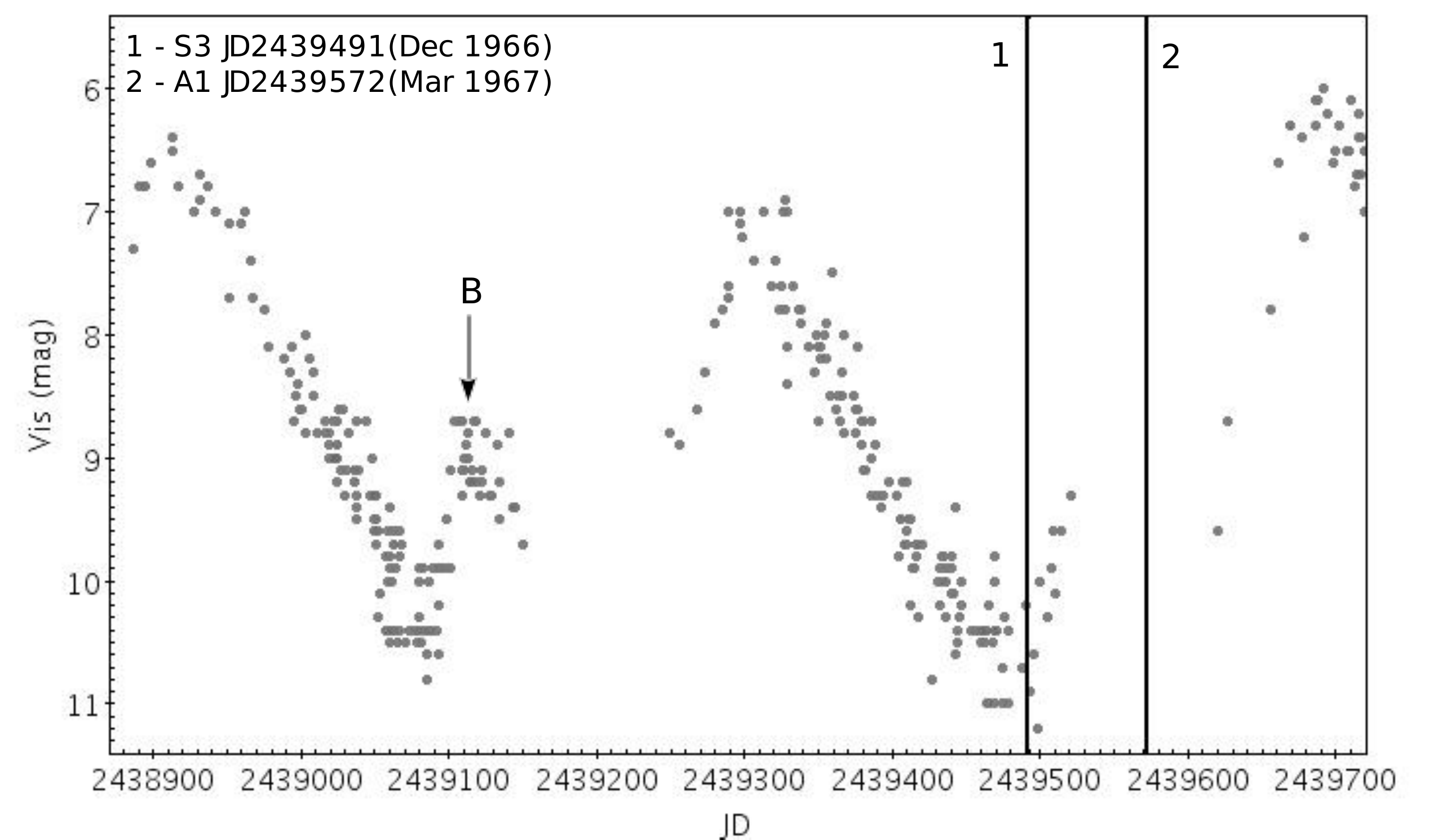}}
\resizebox{\hsize}{!}{\includegraphics{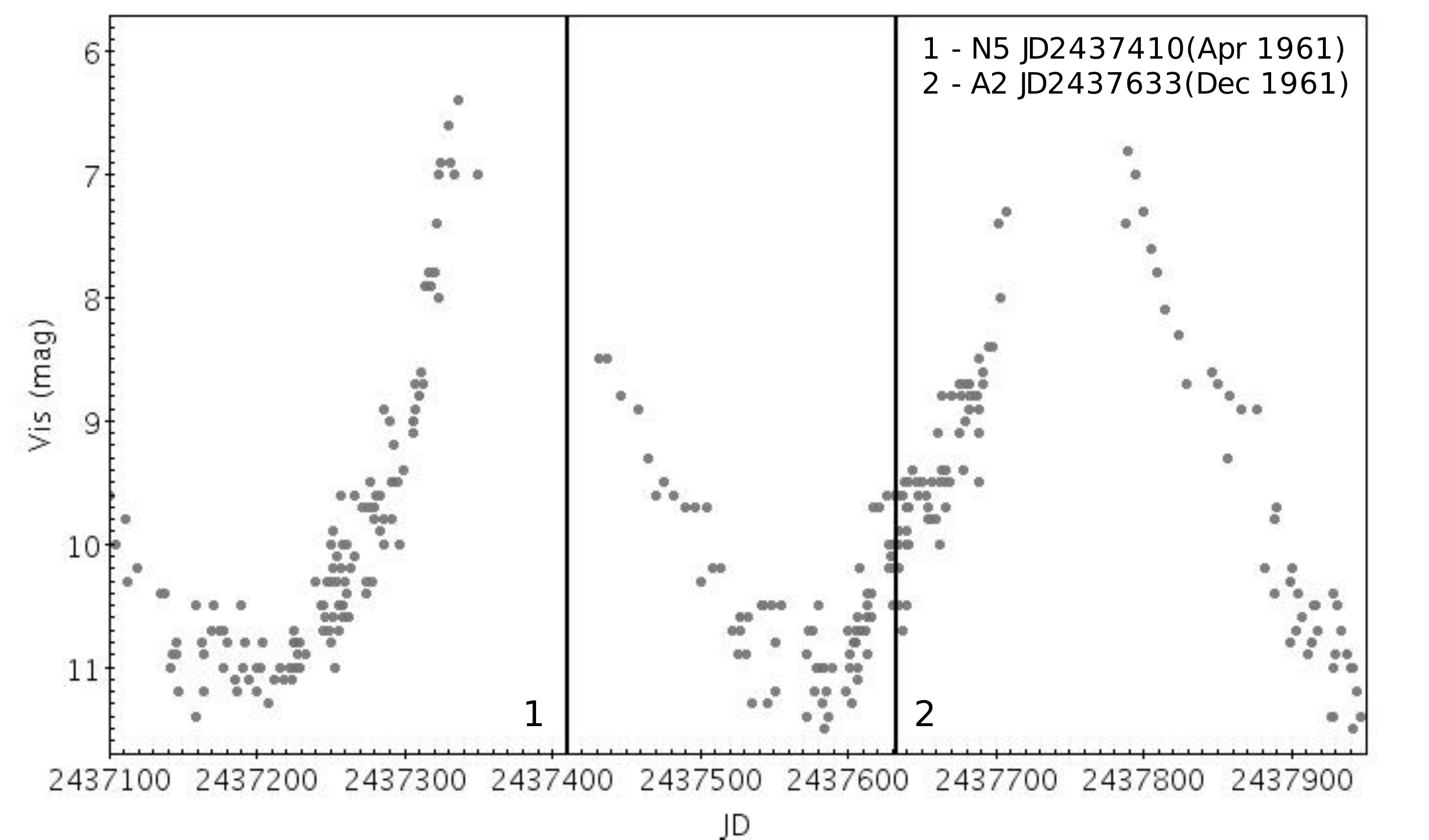}}
\resizebox{\hsize}{!}{\includegraphics{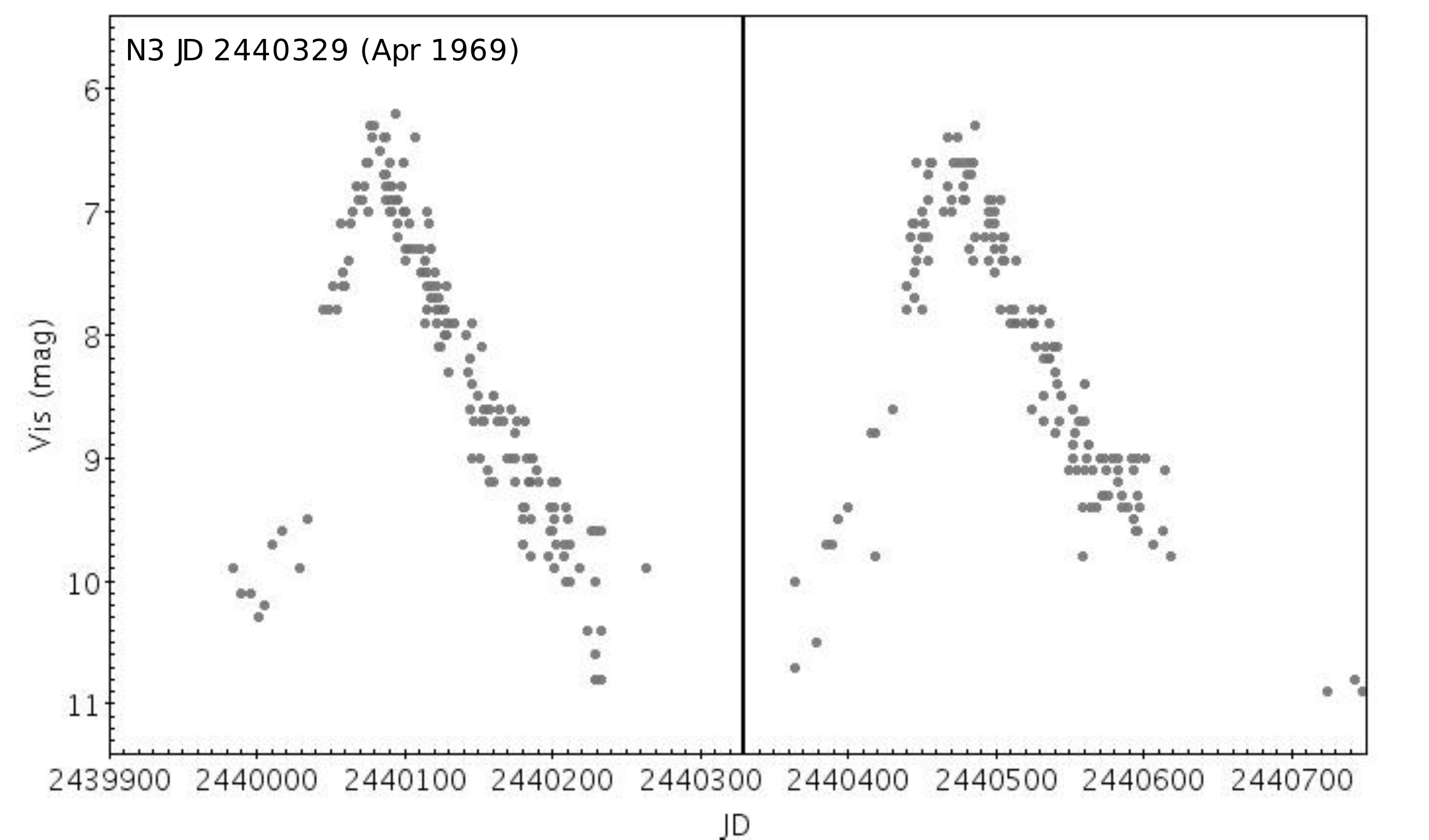}}
\caption{\label{knot_lc} Estimated times of the emission knots ejections. The R~Aqr light curve in visual light is taken from the AAVSO data base. 
The vertical lines mark the estimated ejection epochs of some knots, calculating backwards their kinematical age. The brightening (B), which is 
breaking the harmonic behaviour, occurred after the minimum of the Mira around JD 2439070 (Nov 1966) and is marked by an arrow. Within the measurement 
errors the time of this brightness feature coincides with the extrapolated time of the ejection events of the knots S3 and A1.}
\end{figure}

\begin{table}
\centering
\caption{\label{PM_1y_tab} Proper motion of the bright [OIII] emission knots from 2013/14 HST observations.}
\begin{tabular}{@{\hspace{1mm}}r|@{\hspace{1mm}}r@{\hspace{1mm}}r@{\hspace{1mm}}|r@{\hspace{1mm}}r@{\hspace{1mm}}|r@{\hspace{1mm}}r@{\hspace{1mm}}r}
\hline\hline
Knot    & R.A.(2000) & Dec. (2000) & r$_{jet}$ & PA  & PM$_{R.A.}$  & PM$_{Dec}$  & PM$_{rel}$\\
 $[$OIII$]$ &  \multicolumn{2}{c|}{(epoch=2014.9)}& [\arcsec] & [\degr]  & \multicolumn{3}{c}{[\arcsec yr$^{-1}$]}  \\ 
\hline 
\multicolumn{6}{l}{NE jet}\\
1  &  23:43:50.1 & -15:16:53 & 14.48 & 36.9 & 0.21  &  0.13 &  0.25 \\
2  &  23:43:50.1 & -15:16:55 & 12.90 & 42.6 & 0.17  &  0.16 &  0.23 \\
3  &  23:43:50.0 & -15:16:55 & 12.65 & 38.3 & 0.18  &  0.0  &  0.18 \\ 
\hline
\multicolumn{6}{l}{SW jet}\\
1  &  23:43:48.8 & -15:17:13 & 12.64 & 228.9 & 0.09  &  0.05 &  0.10 \\
2  &  23:43:48.9 & -15:17:14 & 12.81 & 221.6 & 0.12  &  0.13 &  0.18 \\
3  &  23:43:48.8 & -15:17:14 & 14.03 & 227.0 & 0.17  &  0.11 &  0.20 \\
\hline
\end{tabular}
\end{table}
Assuming a constant velocity of the knots, we calculated the kinematical age of each knot and the epochs of ejection of the knots from the system (the 
last column in Table~\ref{PM_tab}). We compared the epochs with the historical light curve available in the AAVSO database. It holds about 20 
000 visual magnitude measurements of R~Aqr covering JD 2394441(Sep 1854)--2457731(Dec 2016). This light curve shows mostly a periodic 
variability caused by the pulsation of the Mira. Therefore, we inspected the light curve to find brightness events which deviate from the harmonic 
behaviour and which may be interpreted as episodic accretion events occurring in the star-``accretion disk``-system. We found one brightness feature 
in the light curve, which is breaking the harmonic variability and which happened close to the calculated ejection times for the two knots A1 and S3, 
which are located in the opposite lobes (Fig.~\ref{knot_lc}). This brightening (marked as B) occurred after the minimum around JD 2439070=Nov 1966. 
The epoch of this minimum does not coincide with the closest periastron passage, which according to \citet{Hin89} probably took place near 1978. 
Taking into account the PM errors, the kinematical age of the brightness features coincides with the extrapolated time of the ejection events of one 
or both of the knots. However, the difference in distances from the jet base for the A1 (r$_{jet}=4\farcs8$) and S3 knots (r$_{jet}=1\farcs5$) is very 
large and thus, it is doubtful that they were ejected during a single event. We have also put here knot N5, whose estimated ejection epoch is similar 
to that of A2. However, as in the case of knots A1 and S3, the spatial separation between knots N5 (r$_{jet}=1\farcs3$) and A2 (r$_{jet}=4\farcs2$) is 
too large for them to have been ejected simultaneously. Furthermore, Fig.~\ref{knot_lc} shows the estimated ejection time of the prominent knot N3 
whose PM is well determined from the 1990--2014 observations. We did not find any brightness features within the errors associated with this knot. 
Considering the dependence of the PMs with distance, we conclude that the assumption of a constant travel velocity is not reliable for the starting 
section of the jet and instead, the acceleration model, where the knot velocity increases with increasing distance, should be adopted.

We also used the 2013/14 HST images to measure the proper motion of some bright arcs with the available 1-year epoch difference. For this goal, we 
selected the three bright knots in the NE lobe located at $\sim14\arcsec$ from the jet source and three knots at a similar distance 
($\sim13\arcsec$) in the opposite SW lobe on the [OIII] image. As for the archival data, we computed the relative PMs using the position of 
the jet source as a reference point. The results (Table~\ref{PM_1y_tab}) show that the relative values of knot PMs are comparable in both directions,  
with an average PM value of $\sim0.23$ yr$^{-1}$ in the NE direction, which is a bit higher than the average in the SW (PM$_{rel}=0.16$ yr$^{-1}$). For 
these emission arcs located at distances $>10\arcsec$, the PM averaged for both directions corresponds to a spatial velocity of about 200--250 km/s 
for the distance and inclination of the jet. One can see that this value of the gas motion is comparable with that of the A1--2 knots located at a 
distance of about 4\arcsec\,from the jet source (Table~\ref{PM_tab}). It implies that the gas acceleration seems to occur at a few arcseconds during 
the gas propagation and at certain distances it decreases or stops. 

\section{Discussion and results.}
\label{discuss}

The high-resolution HST WFC3/UVIS images reveal the fine structure of the R~Aqr jet allowing us to study the outflow morphology showing kinematical features as well as dynamical behaviour with many details. 

\subsection{The structure of inner highly collimated jet.} The HST observations discovered the highly collimated jet component traced to within 
$\sim4\arcsec$ in the NE direction (880 AU at the distance of R~Aqr). It differs from the lower-excitation and wider-angle outflow which is flowing in 
the same direction. The jet is more collimated, with a full opening angle of $\sim4\degr$ whereas the low-excitation outflow has the opening 
angle of $\sim25\degr$ within the first arcseconds from the central source. We should note that previous optical studies 
\citep[e.g.][]{Par94,Burg92}, the radio- \citep[such as][]{Hol93,Mak04}, and X-ray observations \citep{Kel07} were focused on the  analysis of the 
outflow traced by the bright curved knots which we refer to as the wide-angle flows. Therefore, the definition of the "collimated jet" in these 
studies clearly differs from the "highly collimated jet" presented here in our study. We believe that traces of our "highly collimated jet" can be 
seen on the archival HST [OIII] images as a straight emission spike near the N3 knot \citep[see Fig.~\ref{knots_arx} from the study and Fig.2b 
from][]{Par94}.

In the NE direction, we can describe the structure of this highly collimated jet as the bright wiggling continuous outflow which disrupts into several 
emission loops (L1-2) as well as the chain of small faint knots, N7-11 (Fig.~\ref{flux_s} and Fig.~\ref{injet}). This helical pattern which is clearly detected in [OIII], can be traced in [NII] and H$\alpha$ by the chain of small loopy emission knots. These faint knots were not detected in [OI] because the jet is faint in that line, but the initial section of the NE jet in [OI] also has a "wiggling" section which is similar to that visible in the other lines.  Moreover, the positions of the faint knots are also repeating the "wiggling" pattern. It is also interesting to note that the size of the knots are in general smaller than the diameter of the loops, which implies ongoing collimation.

"Wiggling" outflows are often observed among young stellar jets and protostellar molecular outflows \citep{Eis96,Ter99}. \citet{Ter99} investigated  
such binary systems where the accretion disk, from which the jet originates, is inclined to the binary orbital plane. They concluded that the observed 
jet "wiggling" is a consequence of the jet precession caused by tidal interactions in such non-coplanar binary systems. \citet{Nic09} as well as 
\citet{Hol93} suggested that the precession of the accretion disk around the WD may be responsible for the bending of the wide-angle outflow found in 
the previous studies. In analogy to these observations of young stellar jets, we suggest that the "wiggling" that we also find here for the R~Aqr jet 
may result from disk precession as well. We estimated the precession period using the theory from \citet{Ter99} and the binary parameters taken from 
\citet{Grom09} --- $M_h=0.8 M_\odot$ (the mass of the hot WD companion), $M_p/M_h = 1.65$ (where $M_p$ is the mass of the primary companion), $D = 
15.5$ AU (the mean semi-major axis of the system), and $e=0.25$ (eccentricity).  The value of $R_d$ is unknown in our case and we used $R_d=5$ AU 
giving $D/R_d\approx3$ which corresponds to the average value of $2\leq D/R_d\leq 4$ \citep[the range taken from ][]{Ter99}. For this 
calculation, we assumed that the angle $\delta$ between the disk plane and that of the binary orbit is small enough ($<10\degr$) and we adopted $\cos 
\delta = 1$. Using Equation (1) from  \citet{Grom09}, we derived the precession time of $T\approx530$ yr. This value is quite large for the wiggling 
waves that we see. We estimated the projected spatial wavelength $\lambda_{proj}$ of the "wiggling" wave according to 
$\lambda=\lambda_{proj}/\sin\,i$, where $i$ is the angle between the jet symmetry axis and the line of sight, and $T=\lambda/\upsilon$, where 
$\upsilon$ is the jet velocity, from \citet{Grom09}. Using $i=72\degr$ and $\upsilon\sim100$ km/s, we derive $\lambda_{proj}\approx10500$ AU which is 
more than 20 times larger than the projected length of the observed wiggling outflow ($2\arcsec\approx440$ AU). However, we should note that the 
precessing time strongly depends on the $D/R_d$; the $T$ decreases significantly with increasing $R$. It may also be the case that the 
"wiggling" model developed for YSO jets is not fitting for R Aqr which consists of evolved objects, and both the WD and the disk where the jet 
probably forms are much hotter than YSO systems. Furthermore, we cannot exclude that the steady wiggling might be a sequence of dynamical interactions of the 
two collimated flows tilted to each other.

On the other hand, the magneto-centrifugal launching scenario developed for stellar jets \citep[cf][and references therein]{Fer06,Pud07} predicts a 
transport of angular momentum from the accretion disk to the jet by the accretion/ejection `engine'. As a consequence, the gas flow should keep a 
record of rotation at its base during its further outward propagation, at least in the initial section of the flow. Studies focused on jets from young 
stars in star forming regions report indirect evidence of this process. For instance, \citet{Woi05} reported spectral signatures of rotation in the RW 
Aur jet within the first 1\farcs5 (200\,AU) from the central source. Similar signatures of rotation have been discovered in other stellar jets from 
young stars \citep{Bac02a, Cof04, Cof07}. The helical emission pattern can be interpreted as gas rotating around the jet axis during its propagation. 
This is similar to what is derived from the spectroscopy of the young stellar jets. However, the scale of this process for these different objects is 
different --- whereas the young stars exhibit their rotational signatures within 100--200 AU from the jet base, the R~Aqr jet exhibits its helical 
structure within about 900 AU (4\arcsec) in the NE jet lobe.

The apparent morphology of the inner SW jet (counterjet) is slightly different. First of all, it is interesting to note that the dynamics of the bipolar 
gas flows are not trivial and the SW jet does not mirror the NE jet geometry. The simple geometrical model based on the position of the peak emission 
(marked by the circle on Fig.~\ref{flux_s}) and the straight axis of the bipolar jet shows that the initial section of the SW jet is shifted to the 
east with respect to the NE jet. At the same time, the base of the SW flow coincides with the location of the binary components resolved in 
\citet{Sch17}. The distortion of the emission streams is traced in all observed lines and confirmed by the observations obtained in different years. 
On the higher-resolution image \citep[Fig.~6 from ][]{Sch17}, one can see that both counterjets seem to originate in the same region located to the 
east of the binary and along their initial section the gas counterflows follow the curved trajectory. However, the gas in the NE direction then forms 
the straight wiggling beam whereas the SW outflow becomes very distorted and is described as following a "zig-zag" pattern \citep{Sch17}. Moreover, \citet{Nic09} 
found that the NE and SW jets are clearly not coaxial ($i_{NE}\sim103\degr$ whereas $i_{SW}\sim123\degr$). Such misaligned outflows are usually 
believed to originate from a binary in which the circumstellar disk (or disks) is non-coplanar with the binary orbital plane \citep{Ter99}.

The inner SW jet with its opening angle of $\sim11\degr$ is in general less collimated than its NE counterpart and it does not show the helical 
pattern prominently. Instead, in the SW jet there are bright emission clouds which \citet{Sch17} described as bubbles. These patterns, however, can 
also be a projection of emission loops distorted as we can see them at the jet base. Particularly, these loop-like features can be recognised on 
[NII]- and [OI]-images. The flow may impact a stellar wind or ambient dense material which accumulated there from previous ejections \citep{Mak04}. If 
the outflowing gas has a rotational component, it may undergo distortion during its propagation, forming the lobe with a loopy shape and higher 
opening angle. In that case, however, the question arises as to why this interaction takes place only at one side of the system. Comparing directions of the 
collimated jets in the initial section and the wide-angle emitting gas at our high-resolution maps as well as in \citet{Sch17}, we suppose that the 
flows diverge in the NE direction, whereas in the SW they may converge and collide. Therefore, we conclude that in fact both counterjets display 
emission patterns which can be interpreted as the helical motion of the ejected gas, but with different opening angles. If we adopt the simple 
geometrical model based on the location of the highly collimated NE jet, the position angle of the inner bipolar jet propagating in both directions 
is 35\degr.

\subsection{The outer outflow morphology.} On a larger scale, the WFC3/UVIS images display extended emission in all observed emission lines which is 
prominent at distances up to 20\arcsec\, in NE and  30\arcsec\, in the  SW direction (Fig.~\ref{flux_l}). First of all, one can see that this extended 
emission is forming the curved lobes which consist of a chain of bright emission arcs. All arcs in the opposite lobes are curved in opposite 
directions leading to the general view that both lobes are also curved in opposite directions. The curvature of the lobes suggests a counterclockwise 
rotation of the central object expelling gaseous material \citep{Mic88}. The bright [OIII] and H$\alpha$ lines show the complex dynamics of the 
gaseous filaments in the two lobes. For example, the comparison of the [OIII] maps of the SW lobe between 2013 and 2014 reveals a sudden spin-up of a 
gas bubble which led to fast backward motion with respect to surrounding emission features, which continue their slow gradual propagation. Some 
emission knots show a large azimuthal velocity, whereas their radial motion is small; for example, knot 3 in the NE outflow (Table~\ref{PM_1y_tab}). 

The 2013/14 HST images in H$\alpha$ and [NII] illustrate that the lobes are surrounded by extended areas of faint emission (\textit{blue halo} on 
Fig.~\ref{flux_l}) where the emitting gas probably comes from the bright arcs. The complicated dynamics may have led to the fact that the faint 
emitting gas has been moving even to positions forming a plane orthogonal to the outflow axis. The fine structure of the faint emission allows us to 
discern the wider faint arcs which seem to represent an extension of the bright arcs located at the outflow axis. Therefore, we can see that the 
portion of the emitting gas does not tie to the outflow beam, but can freely travel following its own trajectory with substantial radial and azimuthal 
velocity components relatively to the outflow axis, as supposed in \citet{Hol93}. At the same time, the [OIII] line map shows only a narrow halo 
surrounding the lobes which look more collimated than in the lower excitation lines.

The morphology of this outer outflow was also studied at lower resolution in \citet{Solf85}. They found that this outflow is a part of the more 
extended "hourglass-like expanding shell'' whose axis is orthogonal to the flat "oval-shaped" emission nebulosity of $\sim2\arcmin$ extent 
\citep[Fig.~10 in][]{Solf85}. Comparing the spatial differences between the [OIII] peak emission and radio-emission in the curved outflow, 
\citet{Hol93} concluded that its general bend can reflect the jet precession which may follow an accretion disk precessing counterclockwise. Analysing 
O\,IV lines (1032 and 1038 \AA), \citet{Nic09} suggested that the apparent curved trajectory of the observed O\,IV emission in the vicinity of the 
encasing wall of the hourglass-like cocoon \citep{Solf85} for both jet lobes may be an indication of bow shock interaction with the dense nebular 
wall. However, the HST images show that the emission arcs can develop at different distances from the central source along the two jets. Therefore, we 
believe that the arcs might be formed due to the interaction of the fast collimated jet with the slower gas emitting mainly in low-excitation lines.

The important feature of the R~Aqr jet is bright [OIII] emission which is also observable in some YSO jets. However, the high-excitation lines are 
usually relatively faint there and detected in a few bow shocks only \citep[e.g. in \object{HH\,1} and \object{HH\,2},][]{Raga15}. Measurement of 
[OIII]/[OI] along the R~Aqr jet axis shows that this ratio gradually decreases with distance from the jet source implying slow cooling of the ejected 
gas. The measurements of PM of the [OIII] knots showed that the shocks in the beam propagate with a velocity of about 200--250 km/s. According to the 
planar shock model \citep{Har87} such a speed is high enough to produce the bright [OIII] emission. At the same time, comparing the emission lines 
with models \citet{Burg92} concluded that the pre-shock gas in the energetic jet can be partly photo-ionised by the central stellar source. We believe 
that it occurs at the jet base where the density of ionising radiation from the hot star (WD) is especially high.

We also note that there is a discrepancy in the determination of the gas shock velocities in the jet lobes in various studies. \citet{Solf85} found 
from the spectra of several low-excitation optical forbidden lines the moderate gas velocities --- 50--100 km/s. \citet{Con03} found shock velocities 
of 140--150 km/s most consistent with their models, but a higher value of $\sim300$ km/s was not excluded. On the other hand, the studies based on the 
analysis of UV- and X-ray emission determined shock velocities of 250--300 km/s for the high-excitation lines \citep{Mic94,Nic09,Kel07}. Taking into 
account our result of the measured velocity of the [OIII] knots, we suggest that the high velocity components are belonging to the highly collimated 
jet, whereas the slow shocks originate from the wide-angle outflow.

\subsection{The orientation of the subsystems.}  Here we assume that the base of the highly collimated jet coincides with the accretion disk around 
the hot companion (WD). We note that the collimated NE jet can be traced over quite a  long distance --- the ejected knots can be found at the 
distance of $\sim4\arcsec$ ($\sim880$ AU). Moreover, the influence of this jet propagation can be seen at even larger distances on the [OIII]/[OI] and 
[OIII]/H$\alpha$ ratio maps, where we can trace areas of higher ratio values along the jet axis to distances up to 7\farcs5 (1650 AU) from the central 
source. This is a very interesting feature which shows the stability of the jet position and its collimation over long distances. If we adopt the 
velocity of the ejected gas of 250 km s$^{-1}$ (based on the PM of [OIII] features) then the gas will travel a  distance of about $\sim$2000 AU during 
44 years \citep[the probable orbital period][]{Hol97b}. It implies that the jet propagation is continuous and its direction stable in space. 
Therefore, we can also conclude that the main direction of the jet propagation is not affected by the orbital motion of the system components.

Our detailed images show that the axis of the outer lobes does not coincide with the axis of the inner collimated jet. In particular, this effect is 
visible on the [OIII]/[OI] and [OIII]/H$\alpha$ line ratio maps, where the ratios along the collimated NE-jet axis have higher values than in the 
wide-angle emission (Fig.~\ref{ratio_l}). On the small-scale H$\alpha$ and [NII] maps (Fig.~\ref{flux_s}) one can also see that the wide flow of the 
emitting gas is clearly off the inner jet axis. If we consider this outer outflow as a whole, the position angle of its axis is measured to be 
42\degr\, whereas the PA of the inner jet is only 35\degr. On the other hand, the axis position of the outer outflow seems to precess and this process 
is probably linked to the precession of the orbital plane \citep{Hol93}, and hence, the relative positions of the stellar companions. \citet{Hol97b} 
describe this binary as a system containing a Mira red giant with $R_*\sim300\, R_\odot$, $M_*=1.5-2\,M_\odot$, and $P=387$d. The Mira mass-loss rate 
is estimated to be about an order higher than the WD mass accretion rate \citep{Rag08}, therefore the ejected gas can form the continuous component 
that feeds the wide-angle collimated outflow. On the other hand, \citet{Hin89} found that the binary has orbital eccentricity $e=0.6$, which 
is large enough to cause Roche lobe overflow at periastron. Brightness changes in the pulsation cycle of the Mira, which can be explained as the 
mass transfer processes during the periastron passage, occur periodically \citep{Mat79}. This can also be seen in the AAVSO light curve. Therefore, 
the periodical periastron passage acting together with long-term pulsation can led to the time dependent gas ejections. Summarising the information we 
can describe the morphology of the outer outflow as the continuous wide-angle gas outflow with episodic ejections of material happening on a regular 
basis. Finally, we conclude that the wide-angle outflow and highly collimated jet seem to represent two different outflows with different mechanisms 
of ejection, but they can interact during propagation, especially at the initial section.

\citet{Hir10} reported the young molecular protostellar jet in \object{L1448C}, which resembles the R~Aqr jet in some aspects. First of all, the 
L1448C jet exhibits a similar morphology of a two-component outflow --- the highly collimated jet and a wide-opening angle wind. The outflows of both 
objects are revealed in SiO emission, which may originate in the cold collimated stellar wind. \citet{Hir10} conclude that the co-existence of the 
highly collimated jets and the wide-angle shells can be explained by the ``unified {\it X}-wind  model'' in which highly collimated jet components 
correspond to the on-axis density enhancement of the wide-opening angle wind. At the small scales, one can see that the wide-angle R~Aqr outflow is 
also {\it X}-shaped, especially visible on H$\alpha$- and [NII]-maps (Fig.~\ref{flux_s}). Another protostellar jet (\object{IRAS 04166+2706}) that 
exhibits plain bow shocks which have very similar morphology to those of the R~Aqr jet is presented in \citet{Taf17}. However, the evolutionary stage 
of the R~Aqr system and hence, its structure, should have large differences with respect to this protostellar source and therefore, the outflow model 
may need a proper adjustment. 

Another hint on the outflow orientation could be obtained from the orientation of the binary system. \citet{Solf85} suggested that the morphology of 
the two large ejected shells indicates a basic plane of symmetry which should coincide with the orbital plane of the binary system. On the other hand, 
the central axis of the ``hourglass-like'' shell (PA = 355\degr) is clearly tilted to both our inner- and outer outflows. Therefore, the assumption 
that the cone filled by the emission represents the area outlined by the precessing jet is quite reliable. On the other hand, the positions of the 
binary components are still contradictory. \citet{Rag08} calculated the orbit of the white dwarf companion to R~Aqr based on the orbital elements 
published by \citet{Hol97b}. This model predicts that in 2014 the binary should have been aligned "east-west" with the location of the WD to the east of the 
Mira variable. The binary orbit accounts for the observed positions of the SiO masers, whose orientation and extensions agrees well with the PA of 
both the wide-angle outflow and the collimated jet. On the other hand, the binary was resolved in H$\alpha$ in 2014 \citep{Sch17}. Their results agree 
with a position of both components along an "east-west" direction of (PA$_{binary}=270\fdg5$), but the observations show that the WD is located to 
the west of the Mira. Alas, all these studies show that the orientation of the subsystems is complex and the outflow axis as well as the 
snapshot orbital plane can be strongly inclined to the basic orbital orientation.

\subsection{The brightness asymmetry of the outer jet.} On a larger scale, our highly sensitive HST images show the SW lobe to be fainter than the NE 
lobe in all the lines. On the other hand, on a smaller scale the low excitation forbidden lines show the opposite situation --- the NE jet is, in 
general, fainter than its SW counterpart. On the [NII]/[OI] ratio map we can trace the NE jet at the distance up to 1$\farcs$5, whereas in the 
opposite direction it is seen only up to 2$\farcs$5 (Fig.~\ref{ratio_s}). This brightness asymmetry,  as well as the global asymmetry of other 
properties for the counterparts, has also been found in several young stellar jets; for example, in \citet{Mel08,Mel09,Pod11}. In the case of the 
\object{DG Tau jet}, \citet{Pod11} concluded that the observed global asymmetries can be explained in the framework of an MHD disc wind, if the 
latter propagates in an inhomogeneous ambient medium \citep{Fer06}. Several other mechanisms have been invoked for explaining an asymmetric jet --- 
variations in the near-stellar magnetic-field morphology \citep[e.g. in \object{HD 163296},][]{Was06}, warped circumstellar disks \citep[e.g. 
\object{HH\,111},][]{Gom13}, and interactions between a stellar magnetic field and a circumstellar disk magnetic field \citep[e.g.][]{Mat12}. 
Therefore, the phenomenon of the brightness asymmetry in the jet from the evolved object requires further analysis.

\subsection{The origin of the R~Aqr outflows.} 
Summarising our analysis, we conclude that we detect two different outflows which have different morphologies and are probably produced by two 
different mechanisms. \citet{Burg92} considered different mechanisms which can be responsible for the formation of the jet in a symbiotic binary 
system. They calculated a model where the R~Aqr system includes a hot central star with a radiation temperature of $T_\mathrm{hot}\approx40\,000$ K, a 
luminosity of $L_\mathrm{hot}\approx10\, L_\odot$, and a radius $R\geq0.1\, R_\odot$. The cool Mira variable with $R_\mathrm{M}\approx300\, R_\odot$ 
loses material via a low-velocity cool stellar wind, which is partly captured by the hot WD companion, and as a consequence, the captured gas forms the 
accretion disk. The estimated mass-loss rate from the Mira is $M\approx2.7\times10^{-7}M_\odot\, \mathrm{y}^{-1}$ \citep{Burg92} and various model 
calculations show that the WD accretion rate is about an order of magnitude lower \citep{Burg92,Rag08}. This means that a considerable amount of the 
gas can escape from the system. Therefore, we can assume that the wide-angle outflow, which is prominent mainly in the low-excitation emission lines, 
represents the cool stellar wind emanating from the Mira. An important question arises in this case, however, --- what mechanism allows the outflow 
to collimate? \citet{Burg92} suggest that a plausible scenario for the collimation is the propagation of spherically symmetric stellar wind restricted 
by an ambient medium with anisotropic density and pressure. The ambient material is probably concentrated in an orbital plane allowing the gas to 
emerge into polar directions.

Another mechanism considered by \citet{Burg92} is the accretion/ejection scenario which is usually invoked for YSO jets. \citet{Burg92} also argued 
that an eventual magnetic field must be present in order to form a stable disk around the hot accretion source. The magnetic field will then also be 
responsible for ejecting and collimating the outflow as it is in the paradigm used to explain the collimation of the YSO jets. Taking into account our 
results, we assume that the revealed highly collimated jet originates from the "hot star-accreting disk" system and is powered by this mechanism. 
Moreover, taking into account that the escaping wide-angle flow probably interacts with the collimating jet at its base, we can also suggest that the 
collimation engine may influence on the stellar wind flowing in the same direction as well, allowing and helping the collimation of the flow. This 
scenario seems to be especially attractive for the flow lobe in SW direction, where we do not see a clear distinction between the collimated and 
wide-angle outflows.

Previous studies of the R~Aqr outflow noted the simultaneous presence of both the low- and high-excitation emitting gas \citep{Hol90,Mic94,Con03} 
showing different shock velocity components. We think that the two-component structure of the R~Aqr outflow allows us to better understand the 
location of zones with different excitation and velocities. Our results also show evidence that the straight, highly collimated jet propagates with 
higher velocity than the escaping stellar wind which leads to an additional interaction between the flows and enhanced excitation of the 
surrounding material. \citet{Kel07} suggested that the X-ray emission is a likely result of the collision between high-speed outflow and material that 
is already moving outward. This agrees with the structure of the two-component outflow, where the highly collimated jet collides with parts of the 
curved low-speed wide-angle flows. Comparing the observed emission lines with models, \citet{Burg92} concluded that the gas emission in the R~Aqr 
outflow probably represents a combination of shock excitation plus the photo-ionisation by the central stellar source.

\section{Conclusion.}
\label{Conc}

We present new high-resolution HST WFC3/UVIS observations of the R Aquarii jet obtained in 2013/14 using [OIII]$\lambda$5007, [OI]$\lambda$6300, 
[NII]$\lambda$6583, and H$\alpha$ emission lines. The obtained images reveal fine gas structure of the jet at distances of up to a few tens of 
arcseconds from the central region, as well as in an innermost region, within a few arcseconds around its stellar source. The highly sensitive HST 
data allowed us to discern the two gaseous flows in the symbiotic stellar system with different degree of collimation and distinct morphological 
difference. The images also allowed us to map values of several line ratios such as [OIII]$\lambda$5007/[OI]$\lambda$6300 and 
[NII]$\lambda$6583/[OI]$\lambda$6300.

The high-resolution HST images display the energetic highly collimated jet radiating mainly in the [OIII] line. They reveal, for the first time, this 
highly collimated jet component which can be traced to up to $\sim900$ AU in the NE outflow direction. At these scales, this jet is traced by the 
bright continuous outflow which disrupts to a chain of several emission loops and small faint emission knots. We attributed this jet to the 
high-excitation gas powered by the hot central source (WD) surrounded by an accretion disk. 

The wide-opening angle outflow is prominent mainly in the light of the low-excitation emission lines --- [OI]$\lambda$6300, [NII]$\lambda$6583, and 
H$\alpha$. The axis of the wide-angle outflow is not coincident with that of the highly collimated jet, especially at scales of a few arcseconds at 
the jet base. The origin of the gas emission seems to be a cold stellar wind of the mass-losing Mira variable and partly collimated at the flow base.

The inner collimated jet displays a ``wiggling'' pattern which is clearly detected in all observed lines and which is also often observed in jets from 
young stellar binary systems. We cannot conclude on the origin of this wiggling pattern due to uncertain parameters of the binary system and 
circumstellar disk from which the jet probably originates. Further study is necessary.

Our measurements of the proper motions of distinct features in the outflows show that the PM of the knots vary with distance from the central source 
over a wide range of velocities, that is,  50--250 km/s. The propagation of prominent emission knots agrees in general with the scenario of gas 
acceleration within the 2\arcsec\, from the outflow base. At the same time, our PM measurements of distant knots show that at larger distances the gas 
acceleration seems to decrease or stop. 

The highly sensitive HST WFC3/UVIS camera provides high-S/N images which are powerful tools for the study of the jet morphology and can also provide 
detailed information about jet gas conditions. The simultaneous observations of [OIII], [OI], [NII], and [SII] would allow us to measure basic 
parameters of the ionised gas in the R~Aqr outflow and to map the characteristics in detail.

\begin{acknowledgements}
We thank H.M. Schmid, the referee, for his helpful comments on the paper. J.E. and S.M. received support from the Deutsches Zentrum f\"ur Luft- und 
Raumfahrt under grant 50 OR 1504. This research has made use of the SIMBAD database, operated at CDS, Strasbourg, France. The authors also used 
observations from the AAVSO international database.
\end{acknowledgements}

\bibliographystyle{aa}
\bibliography{paper_RAqr}

\end{document}